\pdfoutput=1

\documentclass[nofootinbib,twocolumn,superscriptaddress,aps,preprintnumbers,amsmath,amssymb,prd]{revtex4-1}

\usepackage{mathrsfs}
\usepackage{natbib}
\usepackage{verbatim}
\usepackage{enumerate}
\usepackage{graphicx,bbm,amsbsy,amsfonts,amssymb,amsmath}
\usepackage{bm}
\usepackage{hyperref}
\usepackage{booktabs}
\usepackage{gensymb}
\usepackage{subfigure}
\usepackage{xcolor}
\usepackage{footmisc}
\usepackage{cancel}
\usepackage{blindtext}
\usepackage{ytableau}

\usepackage{slashed}

\usepackage[normalem]{ulem}

\def\2{{(2)}}
\def\1{{(1)}}
\def\0{{(0)}}
\def\m1{{(-1)}}

\let\oldAA\AA
\renewcommand{\AA}{\text{\normalfont\oldAA}}

\hypersetup{
    colorlinks=true,       
    linkcolor=red,          
    citecolor=blue,        
    filecolor=magenta,      
    urlcolor=blue           
}

\newcommand{\be}{\begin{equation}}
\newcommand{\ee}{\end{equation}}
\newcommand{\ea}{\end{eqnarray}}
\newcommand{\ba}{\begin{eqnarray}}
\begin{document}


\title{Naturalness in the Past from a Time-Dependent Supersymmetry Scale}

\author{Tony Gherghetta}
\email{tgher@umn.edu}
\affiliation{School of Physics and Astronomy, University of Minnesota, Minneapolis, Minnesota 55455, USA}

\author{Kunio Kaneta}
\email{kaneta@ed.niigata-u.ac.jp}
\affiliation{Faculty of Education, Niigata University, Niigata 950-2181, Japan}

\author{Wenqi Ke}
\email{wke@umn.edu}
\affiliation{William I. Fine Theoretical Physics Institute, School of Physics and Astronomy,\\
University of Minnesota, Minneapolis, Minnesota 55455, USA}
\affiliation{
IFAE, BIST and Dept.~de Física, Universitat Autònoma de Barcelona, 08193 Bellaterra, Barcelona, Spain}

\begin{abstract}
We consider the possibility that electroweak naturalness may be epoch dependent. In no-scale supergravity, a rolling quintessence modulus can control the supersymmetry-breaking scale, allowing superpartners to be much lighter in the early universe while remaining heavy today. Fifth-force and varying-constant effects can be suppressed via correlated modular weights in the visible-sector Kähler metric and modulus-dependent superpotential Yukawa couplings, while radiative stability of the quintessence potential requires additional tuning of the field-dependent vacuum energy. The resulting effective theory  has distinctive implications for early-universe phase transitions, relic abundances, and gravitational waves.
\end{abstract}

\date{\today}
\maketitle


\section{Introduction}
 
A longstanding motivation for physics beyond the Standard Model is the radiative instability of the electroweak scale. In supersymmetric (SUSY) theories this motivates superpartners near the weak scale, yet present collider bounds have pushed much of the viable supersymmetric spectrum to higher masses. This naturalness argument is usually formulated with an
essentially time-independent supersymmetry-breaking scale throughout cosmological history. In this Letter we explore a different possibility, namely that naturalness may be epoch dependent. If the supersymmetry-breaking scale is controlled by a rolling quintessence field, then the supersymmetric particle spectrum could have been much lighter in the early universe and become heavy only at late times. The present heaviness of superpartners would then be a feature of cosmological timing rather than a time-independent input.

This possibility is difficult to realize in supergravity. In most supergravity quintessence models~\cite{Brax:1999gp,Brax:2006dc,Chiang:2018jdg}, the supersymmetry-breaking and quintessence sectors are separated because a quintessence field that directly participates in supersymmetry breaking generically becomes too massive to roll, and instead behaves as a cosmological constant. In addition, supergravity quintessence models are strongly constrained, and are even subject to no-go results in broad classes of models~\cite{Kolda:1998wq,Brax:2006np}. The supergravity scalar potential and matter Kähler potential naturally communicate the rolling field to the visible sector, producing time-dependent particle masses, gauge couplings, and $\Lambda_{\rm QCD}$, all of which are tightly bounded by fifth-force searches~\cite{Adelberger:2003zx}, equivalence-principle tests~\cite{Damour:1994zq, Damour:2010rp}, limits on varying constants~\cite{Uzan:2002vq}, and big-bang nucleosynthesis (BBN)~\cite{Fields:2019pfx}.

Instead, we consider a no-scale modulus as both the quintessence field and the dominant source of supersymmetry breaking. 
The no-scale structure gives a classically flat direction~\cite{Cremmer:1983bf,Ellis:1984bm}, while small deformations of the superpotential generate a double-exponential potential suitable for quintessence~\cite{Barreiro:1999zs,Rubano:2001su,Sen:2001xu}. The same modulus sets the gravitino mass $m_{3/2}$, so the overall supersymmetry-breaking scale evolves as the quintessence field rolls. For suitable initial conditions, the field undergoes an early kination regime in which the gravitino mass grows rapidly, followed by a slowly varying regime with late-time acceleration and negligible
present variation of particle masses.

The main obstacle is to hide the rolling modulus from precision tests. 
We treat the modular weights in the K\"ahler metrics of the minimal supersymmetric Standard Model (MSSM) fields and the exponents of the modulus-dependent superpotential Yukawa couplings as effective low-energy parameters. Together they determine the modulus dependence of physical Yukawa couplings, the Higgs vacuum expectation value, gauge thresholds, and soft masses. Requiring the light fermion masses, $\alpha_{\rm EM}$, and $\Lambda_{\rm QCD}$ to be independent of the quintessence background to first order around its present value gives algebraic conditions on these parameters. When these conditions are satisfied, the leading scalar charges $d_{m_{u,d,s,e}}$, $d_e$, and $d_g$ vanish, suppressing both universal and composition-dependent fifth forces. This is a highly restrictive requirement, not a generic feature of no-scale compactifications.

Radiative stability imposes a further independent constraint. The Coleman-Weinberg potential from the superpartner spectrum contains field-dependent contributions of order $m_{3/2}^4$, while heavy thresholds can generate terms of order $M^2m_{3/2}^2$ with $M$ being supersymmetric masses. These terms are controlled by supertrace combinations of the visible and heavy-sector spectrum and must be suppressed for the quintessence potential to remain radiatively stable. The field-dependent Higgs vacuum energy requires a similar tuning already at tree level. This requirement is independent of the modular-weight relations that suppress couplings of the rolling field to visible matter. When these conditions are satisfied, a rolling no-scale modulus can control the supersymmetry-breaking scale without immediately violating fifth-force, varying-constant, or cosmological bounds.

With these assumptions, the framework leads to a distinctive cosmological history. The gravitino mass can be of order
$100$~TeV today and many orders of magnitude smaller before BBN. The early universe may therefore have passed through periods in which the MSSM spectrum was light, modifying the electroweak and QCD phase transitions, relic abundances, and possible gravitational-wave signals. Below we construct the no-scale model, derive the modular-weight conditions required by fifth-force constraints, and present illustrative solutions and discuss the resulting phenomenological consequences.


\section{Quintessence and supersymmetry breaking}
\label{sec:model}

\subsection{No-scale supergravity framework}

We decompose the model into a quintessence sector containing a modulus field
$T$, which provides the dominant $F$-term source of SUSY breaking, and a visible
sector containing the MSSM fields:
$K=K_Q+K_{\rm vis}$ and $W=W_Q+W_{\rm vis}$, where $K_{\rm vis}$ contains matter metrics that may depend on $T$.
The quintessence sector is taken to have the no-scale
form~\cite{Cremmer:1983bf,Ellis:1984bm}
\begin{equation}
\begin{aligned}&
    K_Q=-3M_P^2\log \frac{T+\bar{T}}{M_P} ,\\&W_Q=\mu_T^3\left[1  -\epsilon  ( T /M_P)^p-\epsilon^\prime  ( T /M_P)^q\right]\,,
\end{aligned}\label{kqwq}
\end{equation}  
where $\mu_T$ is a mass parameter, $p,q$ are real numbers, $\epsilon,\epsilon^\prime$ are dimensionless parameters satisfying $|\epsilon|,|\epsilon^\prime| \ll 1$ , and the (reduced) Planck scale is $M_P\simeq 2.435\times 10^{18}$~GeV.  
Assuming stabilization of the imaginary part of $T$ at zero with $m_{\rm Im T}\gg H$, without affecting the real direction (see e.g.~\cite{Ellis:2013nxa}),
the canonically normalized real field $\sigma$ is defined by 
\begin{equation}
    \text{Re}\,T=T_0\,e^{\sqrt{\frac{2}{3}}\frac{\sigma}{M_P}}\,,
    \label{eq:sigmadefn}
\end{equation}
where $T_0$ is a model-dependent reference value. At zeroth order in $\epsilon,\epsilon^\prime$, the gravitino mass is
\begin{equation}
    m_{3/2}^2(\sigma) 
    =\frac{\mu_T^6}{M_P} (T+\bar T)^{-3}
    = \frac{\mu_T^6 }{8M_P^4t_0^{3}  }  e^{-\sqrt{6}\frac{\sigma}{M_P}}\,,
    \label{gravmass}
\end{equation}
where $t_0\equiv T_0/M_P$. At this order, $\partial_TW_Q=0$, but the
supergravity covariant derivative
$D_TW_Q=\partial_TW_Q+K_TW_Q/M_P^2=-3W_Q/(T+\bar T)$ is nonzero.
Consequently, the auxiliary field 
$F^T=-e^{\frac{K}{2M_P^2}}K^{T\bar T}D_{\bar T}\bar W_Q$ is nonzero, with 
$|F^T|=(T+\bar T)m_{3/2}$. The modulino shift,
$\delta\chi^T\supset\sqrt{2}\varepsilon F^T\neq0$,
with supersymmetry transformation parameter $\varepsilon$,
implies that the no-scale background spontaneously breaks local
supersymmetry. Nevertheless, the no-scale relation
$K_{T\bar T}|F^T|^2=3m_{3/2}^2M_P^2$ cancels the vacuum energy, leaving
a SUSY-breaking Minkowski background with a flat $\sigma$ direction.

We assume a constant and universal gauge kinetic function for the SM
gauge groups, so the tree-level gaugino masses vanish. Using the leading
no-scale relation $K_TF^T\simeq-3m_{3/2}M_P^2$, the one-loop
anomaly-mediated contribution~\cite{Bagger:1999rd} reduces to
\begin{equation}
m_{\lambda_a}\simeq-\frac{g_a^2}{8\pi^2}
\left[
T_R^a m_{3/2}
+F^T\partial_T\sum_rT_a(r)\log Z_r
\right],
\label{eq:gauginomass}
\end{equation}
where for each SM gauge group with gauge coupling $g_a$, $T_a(r)$ is the Dynkin index of matter field $r$,
$T_R^a\equiv\sum_rT_a(r)$, and $Z_r$ is its diagonal K\"ahler metric.
The scalar soft masses arise at tree level,
$\widetilde m_r^2=m_{3/2}^2-F^{\bar T}F^T
\partial_{\bar T}\partial_T\log Z_r$. For modular metrics, $Z_r\propto(T+\bar T)^{-n_r}$ with modular weights $n_r$ (to be introduced in
Sec.~\ref{sec:modelconstraints}), these
relations give $\widetilde m_r^2=(1-n_r)m_{3/2}^2$ and
$|m_{\lambda_a}|\propto
|\sum_rT_a(r)(1-n_r)|m_{3/2}$.  Thus, in this no-scale background, $m_{3/2}$ sets the overall soft-mass
scale, while the gaugino masses are loop suppressed and generically
nonzero unless the weighted sums vanish.

The $\epsilon,\epsilon^\prime$ deformations lift the no-scale flat
direction at linear order, giving
\begin{equation}
  \begin{aligned}
  V_{\rm Q}(\sigma)
  =&\frac{\mu_T^6}{2M_P^2 t_0^3}\left[ \epsilon p\, t_0^{p}e^{\sqrt{\frac{2}{3}}(p-3)\frac{\sigma}{M_P}}+\epsilon^\prime q\, t_0^{q}e^{\sqrt{\frac{2}{3}}(q-3)\frac{\sigma}{M_P}}\right].
  \end{aligned}
  \label{Vamsb}
\end{equation}
The potential therefore has a double-exponential form, which is known to be a viable quintessence model \cite{Barreiro:1999zs,Rubano:2001su,Sen:2001xu}. Typically, the exponent in $e^{\lambda\frac{\sigma}{M_P}}$ can be chosen to be $\lambda\gtrsim 9$ for radiation to dominate in the early universe, while $0\leq \lambda<\sqrt{2}$ permits late-time acceleration. Since $\epsilon,\epsilon^\prime$ can be arbitrarily small, the overall scale of the potential can be adjusted to reproduce the observed dark-energy density.

Quantum corrections generically shift the field-independent part of the potential, and we assume that this contribution is tuned to zero. Although the superpotential parameters are protected by the non-renormalization theorem, the K\"ahler potential receives corrections that induce field-dependent contributions to $V_Q$~\cite{Brax:1999yv,Hebecker:2019}.  
Even if the light SM particle masses relevant for fifth-force
constraints are chosen to have vanishing first derivatives
with respect to $\sigma$ at its present value
(see Sec.~\ref{sec:modelconstraints}), the superpartner,
electroweak-sector, and heavier-fermion masses remain
$\sigma$ dependent and generate field-dependent
vacuum-energy corrections.  
For the subset of states whose masses scale linearly with the supersymmetry-breaking scale, we may write $m_i=\alpha_i m_{3/2}(\sigma)$, giving the Coleman-Weinberg potential contribution
\begin{equation}
    \Delta V_{CW}=\frac{1}{64\pi^2}\sum_i (-1)^{2s_i}g_i \alpha_i^4  m_{3/2}^4\left[\log \frac{\alpha_i^2 m_{3/2}^2}{\mu^2}-\xi\right]\,,\label{CWpot}
\end{equation}
where $s_i$ is the spin, $g_i$ counts the corresponding degrees of freedom, $\xi=3/2\, (5/6)$ for scalars/fermions (vectors) and $\mu$ is the renormalization scale.
The coefficients $\alpha_i$ are fixed by the particle spectrum, so
cancellation of the terms in Eq.~\eqref{CWpot} requires spectrum-dependent
supertrace relations. Heavy multiplets with $\sigma$-independent
supersymmetric masses $M$ and soft splittings of order $m_{3/2}$ can
additionally generate terms of order $M^2m_{3/2}^2(\sigma)$, while
states with other mass scalings give further $\sigma$-dependent
contributions. We do not construct a complete spectrum realizing these cancellations.
Instead, we assume that the renormalized field-dependent loop and
threshold corrections are tuned sufficiently small to preserve the
value, slope, and curvature of $V_{\rm Q}$. This radiative-stability
tuning, familiar in quintessence models coupled to massive fields,
is distinct from tuning the field-independent cosmological constant.
It could arise from landscape selection of sufficiently
flat potentials~\cite{Garriga:2003hj}, a possibility we do not quantify.

\subsection{Cosmological evolution in the effective theory}

Motivated by the no-scale supergravity form of the potential \eqref{Vamsb}, we next solve for the cosmological evolution of the scalar field $\sigma$ using the Friedmann and scalar equations. It is convenient to use the number of e-folds $N\equiv\ln a$, where $a$ is the scale factor, and denote the derivative $d/dN$ with a prime $(^\prime)$. With this convention, the past corresponds to $N<0~(a<1)$, the present is $N=0~(a=1)$, while the future is $N>0~(a>1)$. The differential equations then become
\begin{equation}
    \begin{aligned}
& \sigma^{\prime \prime }+\left(3 +\frac{h^\prime}{h} \right)\sigma^\prime +\frac{1}{h^2}\partial_\sigma {\widetilde V}_{\rm Q} (\sigma)=0\,,\\
&-2h^\prime h= 4\Omega _{r0}e^{-4N}+3\Omega _{m0}e^{-3N}+h^2 \frac{\sigma^{\prime 2}}{M_P^2}\,,
\end{aligned}
\label{eq:eom}
\end{equation} 
where $h\equiv H/H_0$, $H_0\approx67~{\rm km\,s^{-1}\,Mpc^{-1}}$
\cite{Planck:2018vyg}, and
$\widetilde V_{\rm Q}\equiv V_{\rm Q}/H_0^2$.
Matter is assumed to be decoupled from $\sigma$ in these
reference equations, with the finite-baryon-density force
included in Sec.~\ref{sec:modelconstraints}. The potential is assumed to have the form
\begin{equation}
    {\widetilde V}_{\rm Q}(\sigma)=\widetilde V_0^2\left(e^{\lambda_1\frac{\sigma}{M_P}}+e^{\lambda_2\frac{\sigma}{M_P}}\right)\,, 
\label{eq:simplepot}
\end{equation}
where $\lambda_1=\sqrt{2/3}(p-3)$, $\lambda_2=\sqrt{2/3}(q-3)$. The normalization $\widetilde V_0^2=\rho_{\Lambda 0}/H_0^2\simeq \epsilon p\, t_0^{p-3}\, \mu_T^6/(2H_0^2M_P^2)$ is chosen to match the present-day value of the dark-energy density $\rho_{\Lambda 0}$, assuming $\epsilon^\prime$ is adjusted accordingly. The positive exponents in \eqref{eq:simplepot} drive $\sigma $ towards the runaway direction $-\infty$, while the gravitino mass \eqref{gravmass} increases. The ratio of energy densities is given by
\begin{equation}
  \Omega_{i0} = \frac{\rho_{i0}}{3 M_P^2H_0^2}\,,
\end{equation}
where $\rho_{i0}$ is the present-day energy density for component $i$ with $\Omega_{m0} =0.3149,\,\Omega_{r0} =10^{-4},\,\Omega_{\sigma 0} =0.685$~\cite{Planck:2018vyg} corresponding to the present-day matter, radiation, and dark energy fractions.

\begin{figure}[ht]
\centering
\hspace*{-5mm}
\includegraphics[width=0.48\textwidth]{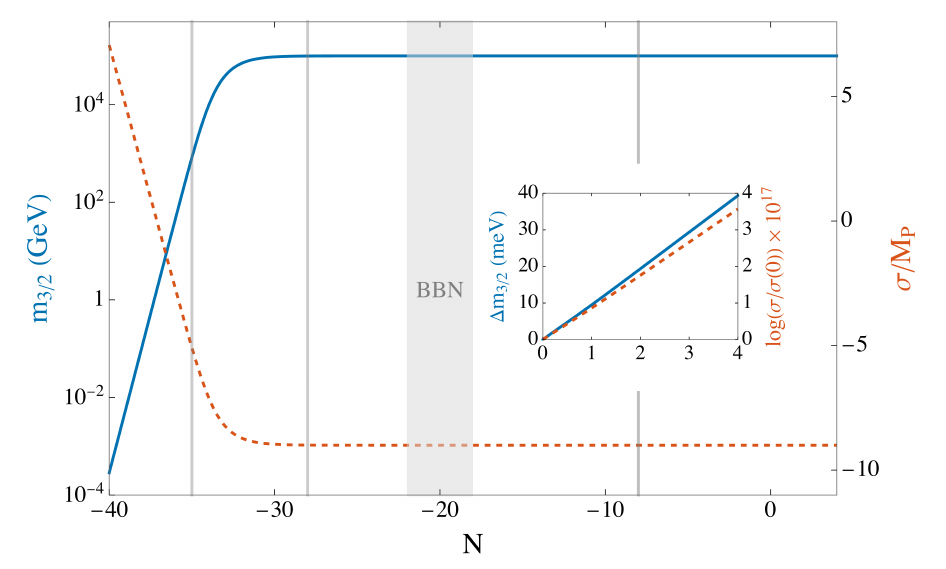}
\caption{The evolutions of $m_{3/2}$  and $\sigma$  with e-fold $N$ for
$m_{3/2,0}=100~{\rm TeV}$. The insets depict  the evolutions near $N=0$.
Gray lines mark the standard-cosmology epochs for
matter--radiation equality, the QCD transition, and the electroweak
transition at $N\simeq-8,-28,-35$, respectively and the band marks BBN
($-22<N<-18$).}
\vspace*{-1.5mm}
\label{m32sigevol}
\end{figure}

\begin{figure}[ht]
\centering
\hspace*{-5mm}
\includegraphics[width=0.46\textwidth]{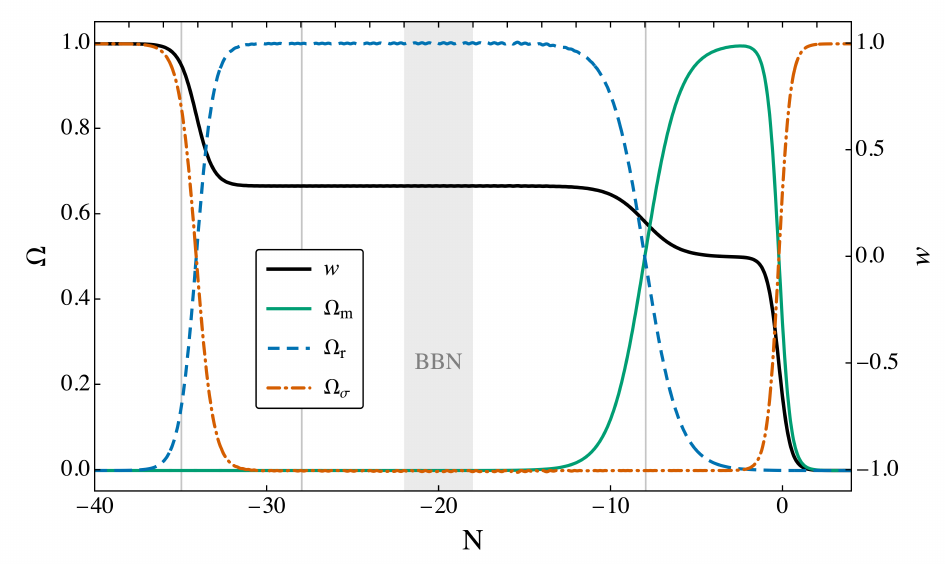}
\caption{The evolutions of the total equation of state $w\equiv p_{tot}/\rho_{tot}$, as well as  the matter (green solid), radiation (dashed) and scalar (dot-dashed) energy-density fractions as a function of the e-folds $N$.}
\vspace*{-1.5mm}
\label{womegaevol}
\end{figure}

To solve \eqref{eq:eom} numerically, we choose $p=15$ and
$q=3+\delta_q$, with $\delta_q=10^{-16}$ and
$\lambda_2=\sqrt{2/3}\,\delta_q$.\footnote{For $q=3$,
the potential \eqref{eq:simplepot} reduces to an exponential
plus a cosmological constant. With suitable initial
conditions, the gravitino mass also grows from smaller
values at early times, but its late-time growth is much
slower than for $q>3$.}
The present-day boundary conditions are
\begin{equation}
\sigma(0)=-9M_P,\qquad
\sigma^\prime(0)=-\lambda_2M_P.
\label{eq:fieldvalues}
\end{equation}
For these boundary conditions, the scalar reaches $\sigma\simeq 7 M_P$ at $N=-40$, with
$\sigma^\prime\simeq-\sqrt{6}M_P$.
The resulting evolution is shown in Figs.~\ref{m32sigevol}-\ref{womegaevol}, neglecting changes in $g_*$, and finite-temperature backreaction. These effects
may modify the dynamics near mass thresholds and shift the electroweak and QCD transitions relative to the standard-cosmology reference epochs indicated in the figures.
Finite-temperature effects are considered in
Appendix~\ref{app:finiteT}, where illustrative solutions retain a substantial increase in $m_{3/2}$ toward the present.\footnote{The quantum fluctuations of $\sigma$ are negligible at later times, while their behavior at earlier times, including the inflationary epoch, depends on the inflationary model~\cite{Malquarti:2002bh}.}

We find that the gravitino mass was significantly smaller in the past, during the early kinetic-energy-dominated phase,
when $\sigma$ rolled rapidly down the potential. 
Consequently, the soft supersymmetry-breaking masses were parametrically smaller in the early universe.
By contrast, near the present epoch, the shallow exponential governs the dynamics, placing the scalar in a slow-roll regime. Its potential energy density dominates with the equation-of-state parameter $w\simeq -1$, driving late-time acceleration. 
To avoid spoiling BBN, we require $m_{3/2}\gtrsim$ 1~GeV by the MeV era, so that even the loop-suppressed gauginos are  generically nonrelativistic and do not contribute appreciably to $\Delta N_{\rm eff}$.

The present day gravitino mass is given by $m_{3/2,0}=\sqrt{\rho_{\Lambda 0}/15\epsilon} \,e^{9\sqrt{6}/2}/(2t_0^{15/2}M_P)$.
For  $m_{3/2,0}=100\text{ TeV}$, we have $\epsilon \simeq 2.5\times 10^{-71}$
and $\mu_T \simeq   9.5\times 10^{11}$~GeV assuming $t_0=0.1$. Extrapolating the bare-potential trajectory gives $m_{3/2}\simeq0.3\text{ MeV}$ at $N=-40$, outside the validated EFT region. 
By contrast, at $N=2$ (approximately 34 billion years from now),
$m_{3/2}$ will only increase by $20\text{ meV}$ relative to its present value.  The future evolution of $\sigma $ is approximately linear in $N$, because $3H \dot{\sigma}\simeq -\partial _\sigma V \simeq -\lambda_2V/M_P$ implies $\sigma^\prime \simeq -\lambda_2 M_P$. Therefore, $m_{3/2}\propto e^{\lambda_2\sqrt{6}N/2}$ continues to grow exponentially, albeit at an extremely slow rate. In particular, for $\lambda_2 =\sqrt{2/3}\times10^{-16}$ the gravitino mass does not reach the Planck scale until $N\sim 3\times 10^{17}$ e-folds, which is exceedingly far into the future.

Finally, let us comment on the parameter choices. The initial conditions in \eqref{eq:fieldvalues} lead to large kinetic energy at very early times. During this kination era, the energy density redshifts as $\rho \simeq \dot{\sigma}^2/2\propto a^{-6}$,  thus $\sigma^\prime \simeq- \sqrt{6}M_P$ and $\sigma$ decreases linearly with $N$, which explains the exponential growth of $m_{3/2}$ in Fig.~\ref{m32sigevol}.  If instead the scalar began with negligible kinetic energy at early times, then $\sigma$ would remain frozen until $H^2\sim V^{\prime\prime}$, and its later evolution $\sigma^\prime \sim -\partial_\sigma V /H^2$ would typically be much slower than linear. Hence, an early kination phase is needed in order to obtain a sufficiently small gravitino mass in the past.

The rolling field $\sigma$ provides an additional, derivative source of
supersymmetry breaking. It induces curvature corrections
$\Delta m_I^2=-n_I\dot{\sigma}^2/(6M_P^2)=\mathcal{O}(H^2)$ 
to visible-sector scalar masses. For our solution, $H/m_{3/2}\lesssim10^{-3}$ from
$N=-40$ onward, so these corrections are negligible over the epochs
considered, but can become important at earlier kination times.

\section{Modular-weight constraints}
\label{sec:modelconstraints}

As noted in section~\ref{sec:model}, supergravity generically couples the quintessence sector to the visible sector and can therefore affect physical observables. To analyze the resulting constraints, we include the visible-sector matter fields $\Phi_I$
via a \textit{twisted} Kähler potential
\begin{equation}
     K_{\rm vis}= \sum_I \frac{M_P^{n_I}|\Phi_I|^2}{(T+\bar{T})^{n_I}}+\frac{c_{ud}\, M_P^{n_\mu} }{(T+\bar{T})^{n_\mu}}(H_u H_d+\text{h.c.})\,,
     \label{visiblek}
\end{equation}
where $I=H_u,H_d,Q,U,D,L,E$ and we have introduced a Giudice-Masiero term~\cite{Giudice:1988yz} with dimensionless parameter, $c_{ud}$. Such moduli-dependent matter metrics are well motivated in string compactifications~\cite{Brignole:1997wnc,Ibanez:1992hc}, where matter fields generically carry modular weights $n_I$. In the effective theory, these modular weights will be treated as phenomenological inputs, specifying the modulus dependence of the visible-sector Kähler metric. We assume that the matter fields relax to a local minimum sufficiently early in the cosmological evolution. Although the Higgs VEVs are much smaller than the Planck scale, they can still contribute  non-negligibly to \eqref{Vamsb}, together with \eqref{CWpot}, and alter the $\sigma $ evolution. Along the real-$T$ trajectory, we assume these corrections are canceled by a suitably tuned perturbation $\delta W_Q (T)$ to the superpotential \eqref{kqwq}, analogous to the field-dependent vacuum-energy tuning in \cite{Hamaguchi:2018vtv}.

The K\"ahler metrics determine the canonical normalization of all MSSM chiral multiplets. At fixed $T$, we denote the canonically normalized
matter fields by
\begin{equation}
    \widehat{\Phi}_I
    =
    \sqrt{K_{\Phi_I\bar{\Phi}_I}}\,\Phi_I,
    \qquad
    K_{\Phi_I\bar{\Phi}_I}
    =
    \left(\frac{M_P}{T+\bar T}\right)^{n_I}.
    \label{eq:cannorm}
\end{equation}
The $c_{ud}$ term in Eq.~\eqref{visiblek} does not generate direct
$H_u$--$H_d$ kinetic mixing, since $K_{H_u\bar H_d}=0$.

The evolution of the quintessence field $\sigma$ induces a corresponding time dependence in the Higgs vacuum expectation value (VEV), $v$~\cite{Brax:2009kd}. To illustrate this effect we restrict to the neutral Higgs sector and write $\widehat H_{u,d}^0=(h_{u,d}+ia_{u,d})/\sqrt{2}$, setting
$a_{u,d}=0$, with $v_{u,d}\equiv\langle h_{u,d}\rangle$ and
$v^2\equiv v_u^2+v_d^2$, where $v_0\simeq246\,{\rm GeV}$ denotes the present value. 
Neglecting Planck-suppressed modulus--Higgs kinetic mixing, the $\sigma$-dependent Higgs potential for
the canonically normalized real fields $h_u$ and $h_d$ is
\begin{equation}
\begin{aligned}
    V\simeq  &\frac{1}{2}\left(1-n_{H_u} 
    +  (n_\mu-1)^2C_{ud}^2\right)m_{3/2}^2  h_u^2\\&
    + \frac{1}{2}\left(1-n_{H_d}
    + (n_\mu-1)^2C_{ud}^2\right)m_{3/2}^2 h_d^2\\
    &
    + 2\left(1-n_\mu \right) \left(1-\Delta-\frac{n_\mu}{2} \right)C_{ud} m_{3/2}^{2}
     h_u h_d \\&
    +\frac{1}{32}(g^2+{g^\prime }^2) (h_u^2-  h_d^2)^2\,,
\end{aligned}
\label{eq:Higgspot}
\end{equation}  
where $C_{ud}\equiv C_{ud,0}\left(\frac{m_{3/2,0}}{m_{3/2}(\sigma)}\right)^{2\Delta/3} $,
$\Delta\equiv (n_{H_u}+n_{H_d})/2-n_\mu$ and the gravitino mass is given in Eq.~\eqref{gravmass}. With this choice, define
$\mu_{\rm eff}\equiv(1-n_\mu)C_{ud}m_{3/2}$.
Additional Higgs quartics induced by the modulus sector
are Planck-suppressed and negligible for the Higgs mass.
After diagonalizing the quadratic potential, we parameterize the lowest Higgs mass eigenvalue as
\be
m_-^2(C_{ud})
=
m_{3/2}^2\left(a_H+f_C(C_{ud})\right)\,,
\ee
where $a_H$ denotes the $C_{ud}$-independent contributions, including possible radiative corrections, while $f_C$ are the $C_{ud}$-dependent contributions. Assuming
$a_H+f_C<0$ and that the Higgs fields adiabatically track their
instantaneous minimum, minimization along the light-Higgs direction gives
\begin{equation}
\frac{v^2}{m_{3/2}^2}
=
-\frac{1}{\lambda_{\rm eff}} (a_H+f_C(C_{ud}))\,,
\label{eq:Higgsvev}
\end{equation}
where $\lambda_{\rm eff}$ is the effective quartic coupling.  In the
minimal no-scale limit, and $n_{H_u}=n_{H_d}=1$, the light
direction is $D$-flat at tree level and
$f_C(C_{ud})=(1-n_\mu)^2C_{ud}^2
-\left|n_\mu(1-n_\mu)\right|C_{ud}$, while $\lambda_{\rm eff}$ includes the required radiative corrections. We require $\Delta>0$ so that $C_{ud}$ grows toward the past,
modifying the Higgs potential.

The evolution of $v$ depends on $a_H$ and $f_C$ and requires
a spectrum-dependent analysis. In this solution, these
contributions nearly cancel today, yielding $v_0\ll m_{3/2,0}$. A superpotential term $W\supset\mu(T)H_uH_d$, with $\mu(T)\propto T^{-(n_{H_u}+n_{H_d})/2}$, generates a contribution $-\mu_0\propto m_{3/2}$.
Our sign convention gives $\mu=-(\mu_0+\mu_{\rm eff})$ and
$b^{(0)}=-2\mu_0m_{3/2}$, with $V\supset-bh_uh_d$.
The diagonal Higgs terms contain $|\mu_0+\mu_{\rm eff}|^2$,
including the cross term.

As $C_{ud}$ grows toward the past, the evolving Higgs potential
can yield $v/m_{3/2}=\mathcal O(1)$ over a finite interval,
reducing the electroweak hierarchy along the local neutral
minimum. At larger $C_{ud}$, the $|\mu|^2\propto C_{ud}^2$ contribution
competes with the $B_\mu\propto C_{ud}$ mixing, leading to electroweak symmetry restoration near
$N\simeq-35.94$. The observed hierarchy today still requires tuning. The equations are given in Appendix~\ref{app:finiteT}.
Figure~\ref{vm32evol} illustrates this behavior using the
zero-temperature one-loop MSSM
hybrid potential \cite{Baer:2012up} with selected SM renormalization group 
improvement, $\mu=-(\mu_0+\mu_{\rm eff})$, and the weak-scale VEV. For $\sigma^\prime(0)=-3\times10^{-12}M_P$, the solution
yields an approximate Higgs pole mass of $125$~GeV today
and the superpartner spectrum shown in Fig.~\ref{mssmspec}. The early VEV excursion remains sensitive to the approximate
particle-threshold treatment. 
Deeper charge- and color-breaking
configurations are present, with the charm-squark channel
giving the smallest estimated present-day tree-level bounce
action among the tested directions, $B\simeq4.8\times10^4$. This estimate suggests
suppressed tunnelling, but a loop-consistent calculation
including other field directions is required to establish
the vacuum lifetime.

\begin{figure}[ht]
\centering
\hspace*{-5mm}
\includegraphics[width=0.45\textwidth]{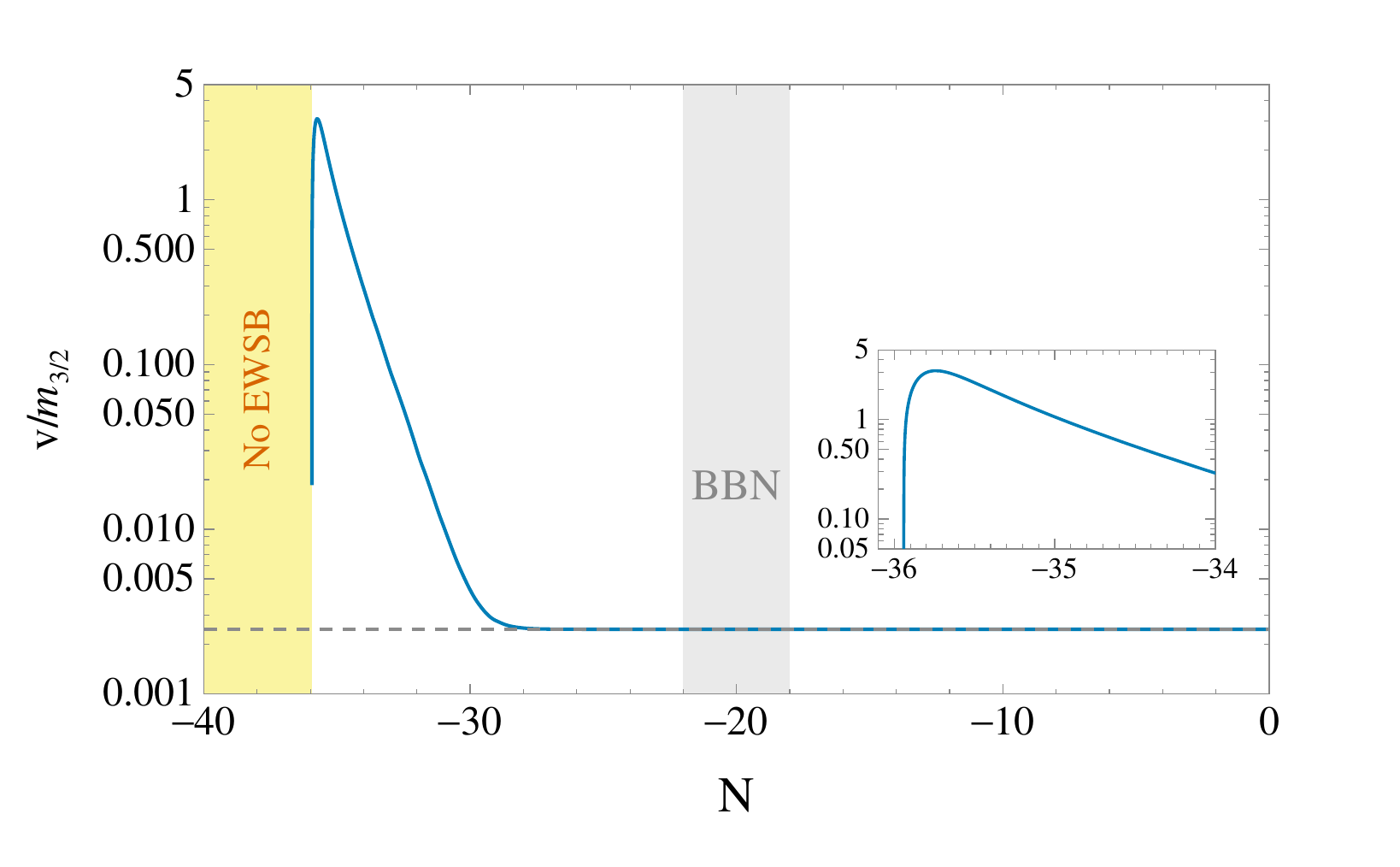}
\caption{The zero-temperature evolution of the matched weak-scale
$v/m_{3/2}$ from the one-loop MSSM hybrid potential
with selected SM RG improvement, assuming $n_{H_u}=9/8$, $n_{H_d}=-1/2$,
$n_\mu\simeq-3.62$ ($\Delta\simeq3.93$),
$\tan\beta_0\simeq4.55$, $m_{3/2,0}=100~{\rm TeV}$,
$C_{ud,0}\simeq0.07$, and
$\mu_0/m_{3/2,0}\simeq0.16$.
The dashed line shows today's ratio,
$v_0/m_{3/2,0}\simeq2.47\times10^{-3}$.}
\vspace*{-1.5mm}
\label{vm32evol}
\end{figure}

The solution in Fig.~\ref{vm32evol} includes the baryonic force
$-\rho_b\partial_\sigma\ln m_B$, described in
Appendix~\ref{app:finiteT}.
The fractional change in the weak-scale VEV between $N=-22$
and today is of order $10^{-6}$, well below the observational
bound $|\Delta v/v|\lesssim0.01$~\cite{Yoo:2002vw}.

Next, direct couplings of the quintessence field to visible matter must be strongly suppressed in order to satisfy present-day fifth-force constraints.  If the physical fermion masses depend on the quintessence background, then fluctuations of $\sigma$ couple to visible matter. Since these fluctuations are ultralight, they mediate a long-range scalar force. Expanding around the present background $\delta\sigma = \sigma-\sigma_0$, we parametrize the relevant low-energy interactions as~\cite{Damour:2010rp}
\begin{equation}
\begin{aligned}
    {\cal L} =&\frac{\delta\sigma}{M_P}\left[d_e \frac{1}{4e^2}F_{\mu\nu}F^{\mu\nu} -d_g\frac{\beta_3}{2 g_3 } G_{\mu\nu}^a G^{a\mu\nu}\right.\\&\left.\qquad -\sum_{i=e,u,d,s}(d_{m_i}+\gamma_{m_i} d_g)m_i \bar{\psi}_i\psi_i\right],
\end{aligned}
\label{eq:bgL}
\end{equation} 
where $F_{\mu\nu}$, $G_{\mu\nu}^a$ denote the electromagnetic and gluonic field strengths, respectively, $\beta_3$ is the QCD $\beta$ function, and $\gamma_{m_i}$ is the mass anomalous dimension. The normalization of the gluonic term is chosen so that $d_g$ couples to the RG-invariant QCD trace anomaly or equivalently measures the scalar dependence of $\Lambda_{\rm QCD}$. In this parametrization, the scalar coupling to a macroscopic body $A$ can be written as $\alpha_A=d_g+\bar{\alpha}_A$ where $d_g$ is the universal QCD contribution and $\bar{\alpha}_A$ encodes composition-dependent effects.
The Cassini measurement of the frequency shift induced by the solar Shapiro time delay~\cite{Bertotti:2003rm} constrains the universal scalar coupling at the level $|d_g|\lesssim10^{-3}$, whereas the non-universal contribution $\bar{\alpha}_A$ is probed by tests of the weak equivalence principle.  Using the parametrization of Ref.\cite{Damour:2010rp},  
the observational bound \cite{MICROSCOPE:2022doy} translates 
into constraints on the relevant composition-dependent combinations of scalar couplings~\cite{Berge:2017ovy}, typically corresponding to 
$d_g\lesssim 10^{-7}$ for generic non-universal couplings.
In the following, we show that the leading couplings of $\sigma$ to visible matter can be suppressed by specific modular-weight relations.

In supergravity, the charged-lepton Yukawa interaction takes the form (see also \cite{Brax:2017bcp,Brax:2023nvx})
\begin{equation}
y_{e}(T)e^{\frac{K}{2M_P^2}}H_dLE
\simeq
y_{e}(T)e^{\frac{K_Q}{2M_P^2}}H_dLE
\equiv
Y_{e}\widehat H_d\widehat L\widehat E\,,
\label{ELmass}
\end{equation}
where $y_{e}(T)$ is the holomorphic superpotential coupling,
$Y_{e}$ is the physical Yukawa coupling, and the hatted fields are
canonically normalized as in Eq.~\eqref{eq:cannorm}. The contribution from $K_{\rm vis}$ to the prefactor $e^{K/(2M_P^2)}$ is negligible, since it is strongly suppressed by small visible-sector field values relative to $M_P^2$. 
Similar expressions also apply to the up- and down-type Yukawa operators
$H_uQU$ and $H_dQD$.  Hence, for any charged fermion $f$, the corresponding physical Yukawa coupling is 
\begin{equation}
Y_f(\sigma)
=
y_f(T)
\left(\frac{T+\bar T}{M_P}\right)^{(N_f-3)/2}.
\label{eq:physicalYukawa}
\end{equation}
where $N_f$ is the sum of the three modular weights appearing in the
corresponding Yukawa operator e.g.
$N_{e_i}=n_{H_d}+n_{L_i}+n_{E_i}$, with analogous definitions for
$N_{u_i,d_i}$. We parameterize the holomorphic modulus dependence as
$y_f(T)=y_{f,0}(T/T_0)^{\kappa_f}$, with approximately constant
$\kappa_f=T\partial_T\ln y_f$, giving
$Y_f\propto m_{3/2}^{\,1-(N_f+2\kappa_f)/3}$.
Assuming the Higgs VEVs have a general $m_{3/2}$ dependence, we define
their present logarithmic derivatives
$x_{f,0}\equiv(d\ln v_{u,d}/d\ln m_{3/2})_0$.
At leading order, the corresponding local mass derivative is
$\xi_f\equiv(d\ln m_f/d\ln m_{3/2})_0
=1+x_{f,0}-(N_f+2\kappa_f)/3$, and the present scalar charge,
$d_{m_f,0}\equiv
M_P(\partial_\sigma\ln m_f)_{\sigma_0}$, becomes
\begin{equation}
d_{m_f,0}
=
\frac{1}{\sqrt6}(N_f+2\kappa_f-3(1+x_{f,0}))
=
-\sqrt{\frac32}\,\xi_f.
\label{eq:fermioncharge}
\end{equation}
In the low-energy parametrization of Ref.~\cite{Damour:2010rp}, the
composition-dependent coupling $\bar{\alpha}_A$ is controlled primarily
by $d_{m_{u,d,e}}$,
$d_g$, and the electromagnetic
coupling $d_e$, with a possible additional strange-quark contribution.
A simple sufficient condition for evading the fifth-force constraints is therefore\footnote{For order-one nonuniversal matter charges, the Planck-scale
field excursion violates the thin-shell condition, preventing chameleon screening~\cite{Brax:2006np} in this no-scale potential.}
$d_{m_{u,d,s,e},0}=d_{g,0}=d_{e,0}=0$.
The light-fermion conditions require
$N_f+2\kappa_f=3(1+x_{f,0})$ for $f=u,d,s,e$.
The heavier-fermion charges need not vanish, but their threshold
contributions must be included in $d_g$ and $d_e$.

For the illustrative $x_{u,0}=x_{d,0}=1$ solution below,
the zero-charge light Yukawa couplings scale as $m_{3/2}^{-1}$.
More general finite present-day VEV derivatives can be accommodated
in the light-fermion charge conditions by shifting
$\kappa_f\rightarrow\kappa_f+\frac32(x_{f,0}-1)$, keeping the
K\"ahler modular weights fixed. The heavy-fermion exponents can
then be adjusted to satisfy the corresponding gauge-threshold
conditions, while the perturbativity boundary remains solution
dependent. Thus charge cancellation does not require a constant
Higgs-to-gravitino mass ratio throughout cosmological evolution. 
These shifts leave the tree-level diagonal scalar soft masses
$\widetilde m_I^2=(1-n_I)m_{3/2}^2$ unchanged,
since they depend only on the K\"ahler modular weights.\footnote{
The trilinear terms
$V_{\rm soft}\supset A_fY_f\phi_i\phi_j\phi_k+\mathrm{h.c.}$,
with $A_f=(N_f+2\kappa_f-3)m_{3/2}$ up to a sign convention,
do depend on these shifts. After electroweak symmetry breaking,
they induce $(M_{\widetilde f}^2)_{LR}\sim m_fm_{3/2}$, which can be subleading while the trilinears still affect CCB minima.}

The gauge-sector charges $d_e$ and $d_g$ are computed in
Appendix~\ref{app:matching}. Assuming the superpartner,
heavy-Higgs and $W$ masses are locally proportional to
$m_{3/2}$, the QED charge below the electron threshold is
\begin{equation}
\begin{aligned}
d_e(\mu<m_e)\propto{}&
-66+S_e+3(n_{H_u}+n_{H_d})
\\
& +8(\xi_u+\xi_c+\xi_t)+2(\xi_d+\xi_s+\xi_b)
\\& +6(\xi_e+\xi_\mu+\xi_\tau)\,,
\end{aligned}
\label{eq:decondition}
\end{equation}
where $S_e$ is defined below Eq.~\eqref{app:beuv}.
For QCD, we retain $u,d,s$ in the low-energy three-flavor theory and integrate out the charm, bottom, and top quarks together with the colored superpartners. Under the same threshold assumptions, this gives
\begin{equation}
    d_g(\Lambda_{\rm QCD})
   =-\frac{1}{9\sqrt{6}}\left(
    3+S_g+2(\xi_c+\xi_b+\xi_t)\right)\,,
    \label{eqn:dg}
\end{equation}
where $S_g$ is defined in \eqref{app:sgdef}.

Together with $d_{m_{u,d,s,e}}=0$, a simple sufficient condition for
evading the observational constraints is
$d_e=d_g(\Lambda_{\rm QCD})=0$. An illustrative solution uses the  integer-valued K\"ahler modular weights
$n_{H_u}=1$, $n_{H_d}=0$,
$(n_{Q_i},n_{U_i},n_{D_i})=(-2,0,-1)$ for $i=1,2$, $(n_{Q_3},n_{U_3},n_{D_3})=(0,0,-1)$, and
$(n_{L_i},n_{E_i})=(0,0)$ for $i=1,2,3$, together with the holomorphic
exponents $\kappa_u=7/2$, $\kappa_{d,s}=9/2$, $\kappa_e=3$,
$\kappa_{c }=1/2$,
$\kappa_{t}=1 $, $\kappa_b=2$, and
$\kappa_{\mu,\tau}=-9/2$. These give
$\xi_{u,d,s,e}=0$, $\xi_c=2$, $\xi_{t,b}=1$, and
$\xi_{\mu,\tau}=5$. For these weights, $S_g=-11$ and $S_e=-23$, and the one-loop
cancellation conditions are
$\xi_c+\xi_t+\xi_b=4$ and
$\xi_c+\xi_t+\xi_\mu+\xi_\tau=13$. Hence $d_{m_{u,d,s,e}}=0$ at leading order and
$d_e=d_g=0$ at one loop. The analytic example assumes $x_{u,0}=x_{d,0}=1$, while
Fig.~\ref{vm32evol} uses Yukawa exponents adjusted for the
calculated variation of the Higgs VEVs and particle masses. All sfermion soft masses are strictly positive at the mediation scale,
while $\widetilde m_{H_u}^2=0$ and
$\widetilde m_{H_d}^2=m_{3/2}^2$.
Choosing $n_\mu<1/2$ gives $\Delta=1/2-n_\mu>0$,
while the unequal Higgs weights generically prevent
alignment of the light Higgs eigenstate with the
$D$-flat direction. The MSSM solution in Fig.~\ref{vm32evol} instead uses
the fractional modular weights given in
Appendix~\ref{app:finiteT}, with Yukawa exponents adjusted
for the calculated Higgs and threshold responses. Thus, suppressing both universal
and composition-dependent fifth forces selects correlated
K\"ahler modular weights and holomorphic Yukawa exponents.

Alternatively, rather than setting the low-energy matter charges to
zero, one may impose the universal relation
$d_g=d_{m_{u,d,s,e}}$, while keeping $d_e=0$. The resulting bound is
the weaker universal-coupling constraint
$|d_g|\lesssim10^{-3}$~\cite{Choi:2018rze}. Treating the $\kappa_f$ as
continuous effective logarithmic exponents, a nearby one-loop universal
solution is obtained from the preceding vanishing-charge solution by
keeping the integer-valued K\"ahler modular weights fixed and taking
$|\delta|\lesssim10^{-3}$,
$\kappa_u=7/2+\delta/2$, $\kappa_{d,s}=9/2+\delta/2$,
$\kappa_e=3+\delta/2$, $\kappa_c=1/2$, $\kappa_t=1$,
$\kappa_b=2+27\delta/4$, and
$\kappa_{\mu,\tau}=-9/2-15\delta/8$. The corresponding mass exponents
are $\xi_{u,d,s,e}=-\delta/3$, $\xi_c=2$, $\xi_t=1$,
$\xi_b=1-9\delta/2$, and
$\xi_\mu=\xi_\tau=5+5\delta/4$. Including the $c,b,t$ thresholds in
$d_g$ and all charged-fermion thresholds in $d_e$, these exponents give
$d_g=d_{m_{u,d,s,e}}=\delta/\sqrt6$ and $d_e=0$ at leading order for the light-fermion charges and one loop for the gauge charges, neglecting hadronic corrections. At
$\delta=0$, the solution reduces to the vanishing-charge solution,
while the K\"ahler modular weights and tree-level scalar masses remain
unchanged.

Both the vanishing-charge and universal solutions are one-loop matching
relations. Below all colored-superpartner and heavy-quark thresholds, one
may write schematically $d_g=d_g^{(1)}+\Delta_g^{(2)}+\cdots$, where
$d_g^{(1)}$ includes all one-loop threshold contributions. The two-loop correction
to an individual QCD-threshold contribution is suppressed by
${\cal O}(\alpha_s/\pi)$ relative to its one-loop value, whereas the net
shift $\Delta_g^{(2)}$ is spectrum dependent and analogous higher-order
corrections affect $d_e$.
Any additional heavy sector invoked to cancel radiative corrections
to the quintessence potential may also modify the matching
conditions, potentially already at one loop, if it couples to
the corresponding gauge sectors.
We assume that the charge conditions can be restored by adjusting
the effective modular weights and/or holomorphic exponents.
Verifying this assumption requires the full spectrum and its
$\sigma$ dependence.

 
\section{Cosmological consequences}
\label{sec:pheno}

The most robust consequence is a time-dependent particle spectrum.
Since the soft masses track $m_{3/2}(\sigma)$, almost all superpartners can be
much lighter in the early universe than today. Matching the bare-potential solution at $N=-40$ to an earlier
inflationary or radiation-dominated phase gives approximately
six e-folds of kination, consistent with the approximate
ten-e-fold bound of Ref.~\cite{Eroncel:2025bcb}.
Finite-temperature matter effects can produce damped oscillations
and become negligible as the plasma cools. The baryonic force
persists and can dominate the late-time scalar motion, while
the quintessence potential supplies the dark-energy density. The MSSM solution in Fig.~\ref{vm32evol} becomes
electroweak symmetric near $N\simeq-35.94$ and extends
to $N=-40$, as do the singlet and thermal solutions
in Appendices~\ref{app:nmssm_ccb} and~\ref{app:finiteT}.

The evolving spectrum also changes the gauge and Yukawa couplings. At $N=-35$, the one-loop relation gives
$\Delta\alpha_3^{-1}(10^{16}\,{\rm GeV})\simeq-0.14$
for Fig.~\ref{vm32evol}, compared with $+0.089$ and $-0.26$
for the thermal MSSM and NMSSM solutions, respectively.
The electroweak gauge couplings remain perturbative. Earlier extrapolation
requires nonperturbative visible-sector dynamics. As $m_{3/2}$ decreases, all MSSM Yukawa couplings decrease
except $Y_t$ and $Y_b$, which remain constant.
All NMSSM Yukawa couplings decrease, with weak variation
in $Y_d$ and $Y_s$. Most MSSM couplings decrease much faster because their powers of $m_{3/2}$ are larger.
Thermal oscillations modify these trends.
These changes affect interaction rates, thermal masses,
particle thresholds, and two-loop gauge evolution.
Neutrino masses and their force on $\sigma$ require a separate study of the neutrino mass-generation mechanism.

These effects open several possible cosmological signatures. If
kination overlaps the QCD epoch, the altered expansion rate and
evolving strong scale can shift axion oscillations and enhance
the misalignment abundance for fixed
$f_a$~\cite{Visinelli:2009kt}. In variants where electroweak
symmetry remains unbroken until confinement, a QCD condensate
could induce Higgs tadpoles and initiate electroweak symmetry
breaking. The system would then approach its late-time neutral
vacuum. Consistency with BBN requires the weak scale, gauge
couplings, and strong scale to approach their standard values
sufficiently early, for example by $T\sim10$~MeV.

In variants with six light quarks and light superpartners near
confinement, the plasma differs from ordinary $2+1$-flavor QCD.
The chiral-limit analysis for $N_f\geq3$ suggests a first-order
transition~\cite{Pisarski:1983ms}, but light superpartners require
a dedicated finite-temperature calculation. A sufficiently strong transition could produce a
stochastic gravitational-wave background, with a
nanohertz-to-microhertz peak for a transition near the GeV scale and
higher frequencies for an earlier transition.

Dark matter is model dependent. An axion or another dark-sector
candidate is possible,
whereas a neutralino-like state whose mass tracks $m_{3/2}$
carries an order-one scalar charge and requires a coupled
dark-matter--quintessence analysis if appreciably populated. The displayed histories assume negligible supersymmetric relic
density and scalar force, potentially through $R$-parity-violating
decays. This assumption and the axion's modulus dependence require
a separate relic analysis. Quantitative predictions also require
a more complete treatment of the thermal spectrum and QCD
thermodynamics. Nevertheless, the solutions provide an explicit
effective-theory illustration of an evolving supersymmetry scale
and late-time acceleration including thermal and baryonic
backreaction, subject to the required vacuum-energy tuning
and negligible dark-matter coupling to quintessence.

\section{Conclusion}
\label{sec:conclusion}

In this Letter we have presented a no-scale supergravity framework in which a rolling quintessence modulus controls the supersymmetry-breaking scale. The central consequence is that the superpartner spectrum can be lighter in the early universe and heavy today, while modular-weight and modulus-dependent Yukawa coupling parameters are chosen to suppress the leading fifth-force and varying-constant effects. Within the stated approximation, a singlet solution realizes these conditions with positive matter-field soft masses,
an approximate $125$~GeV Higgs, and no lower CCB
minimum found in our search. Thermal effects become negligible as the plasma cools,
while the baryonic force can continue to control the scalar motion and the quintessence potential supplies the dark-energy density.
This opens unconventional possibilities for early-universe dynamics, including modified relic production and altered cosmological phase transitions.

Several important questions remain open. The field-dependent Higgs vacuum energy and radiative corrections require an additional tuning involving the full spectrum, including hidden and heavy thresholds, whose microscopic origin remains unspecified. Higher-order and heavy-threshold corrections may also shift the one-loop charge conditions. Although our solutions include one-loop thermal effects,
selected higher-order corrections and changes in the
relativistic degrees of freedom, quantitative predictions for relic production, the electroweak and QCD
transitions, gravitational waves, and BBN require a more complete cosmological analysis. Further extensions include opposite modulus rolling, realizing
``naturalness in the future'', and gauge-, anomaly-, or
hybrid-mediated supersymmetry breaking.
Generalized no-scale K\"ahler potentials and alternative leading
superpotentials could also change the relation between modular
weights and soft masses. A complete UV embedding remains an
important goal, but these effective-theory examples show how
the electroweak hierarchy can evolve while leading present-day
scalar couplings remain suppressed.

\medskip\noindent\textit{Acknowledgments\,---\,}%
We thank Philippe Brax for helpful comments on the manuscript. T.G. was supported in part by the Department of Energy under Grant No. DE-SC0011842 at the University of Minnesota; K.K. by the Niigata University Grant for the Enhancement of International Collaborative Research; and W.K. by the European Union’s Horizon 2025 research and innovation program under Marie Skłodowska-Curie Grant Agreement No. 101270027. OpenAI Codex assisted with constructing the numerical solutions,
manuscript editing, and consistency checks. 
K.K. and W.K. thank Institut Pascal at Université Paris-Saclay, where part of this work was completed, with support from “Investissements d’avenir” ANR-11-IDEX-0003-01, the P2I axis of the university’s Graduate School of Physics, the CNRS-IRP project UCMN, IJCLab, CEA, and ACME.

\appendix

\section{One-loop matching of the gauge-sector scalar charges}
\label{app:matching}

In this Appendix, we calculate the radiatively generated
gauge-sector charges $d_e$ and $d_g$ used in Sec.~\ref{sec:modelconstraints}. A constant gauge kinetic
function gives no tree-level $\sigma$ coupling to gauge fields.
The relevant $\sigma$-dependence follows
from the expression for the physical running gauge coupling \cite{Shifman:1986zi,Kaplunovsky:1994fg}. At a fixed physical scale $\mu$, its one-loop $\sigma$-dependent part is
\begin{equation}
\begin{aligned}
\left.\frac{1}{g_a^2(\mu)}\right|_{\sigma}
=&\,
\frac{c_a}{16\pi^2}\frac{K_Q}{M_P^2}
-\sum_r \frac{T_a(r)}{8\pi^2}\log\det Z_r 
\\
&+\left.\frac{b_a}{16\pi^2}
\log\frac{\Lambda^2}{\mu^2}\right|_{\rm thresholds}\,,
\end{aligned}
\label{eq:KLsigma}
\end{equation}
where $b_a=\sum_r N_rT_a(r)-3 T_G^a$, and
$c_a=\sum_r N_rT_a(r)-T_G^a$, with $N_r$ denoting multiplicities. The last
term with UV cutoff scale $\Lambda$ is $\sigma$-independent within a fixed effective theory, but changes when
fields with $\sigma$-dependent masses are integrated out (such as the electroweak-charged superpartners and Higgs field via \eqref{eq:Higgsvev}). Define
$s\equiv\sigma/M_P$ and the local mass slope
\begin{equation}
 p_j\equiv
 \left.\frac{d\ln m_j}{d\ln m_{3/2}}\right|_0,
 \qquad
 \partial_s\ln m_{3/2}=-\sqrt{\frac32}.
 \label{app:massslope}
\end{equation}
For a charged fermion, $p_f=\xi_f$. Differentiating the physical gauge-coupling relation \eqref{eq:KLsigma} gives, above the particle thresholds,
\begin{equation}
 \left(\partial_s g_a^{-2}\right)_{\rm UV}
 =\frac{\sqrt{2/3}}{16\pi^2}
 \left[-3c_a+2\sum_I T_a(I)n_I\right].
 \label{app:uvmatching}
\end{equation}
The sum includes representation multiplicities. If a state of physical
mass $m_j$ is integrated out, let
$\Delta b_{a,j}=b_{a,<}-b_{a,>}$. Continuity of the coupling at $m_j$
then gives
\begin{equation}
 \Delta\!\left(\partial_s g_a^{-2}\right)
 =-\frac{\sqrt6}{16\pi^2}\,\Delta b_{a,j}p_j.
 \label{app:thresholdrule}
\end{equation}
Equations~\eqref{app:uvmatching} and \eqref{app:thresholdrule} are the only
ingredients needed below. For mixed or nondegenerate states,
Eq.~\eqref{app:thresholdrule} is applied to each physical eigenmass with its own local slope.

\subsection{Electromagnetic charge}

With GUT-normalized hypercharge,
$e^{-2}=\frac53g_1^{-2}+g_2^{-2}$, we define
\begin{equation}
 (\partial_s e^{-2})_0
 \equiv\frac{\sqrt{2/3}}{24\pi^2}B_e.
 \label{app:edefs}
\end{equation}
The K\"ahler and super-Weyl terms in
Eq.~\eqref{app:uvmatching} give
\begin{equation}
 B_e^{\rm UV}=-72+S_e+3(n_{H_u}+n_{H_d}),
 \label{app:beuv}
\end{equation}
where $S_e\equiv\sum_{i=1}^3
\left(5n_{Q_i}+4n_{U_i}+n_{D_i}
 +3n_{L_i}+3n_{E_i}\right)$.
For masses proportional to $m_{3/2}$ near
today's field value, the sfermion, Higgsino, wino,
and heavy-Higgs thresholds give
$\Delta b_{\rm em}=-25/3$ and $\Delta B_e=75/2$.
The massive-$W^\pm$ threshold, including Goldstone and ghost
contributions, gives $\Delta b_{\rm em}=7$ and
$\Delta B_e=-63/2$. Together these change the constant
term from $-72$ to $-66$.
A charged Dirac fermion gives
$\Delta b_{{\rm em},f}=-4N_c^fQ_f^2/3$. Equation~\eqref{app:thresholdrule}
therefore adds
\begin{equation}
 \Delta B_e^f=6N_c^fQ_f^2\xi_f,
 \label{app:fermionweight}
\end{equation}
corresponding to weights $8$, $2$ and $6$ for an up-type quark, a
down-type quark and a charged lepton. Combining these pieces gives
\begin{align}
B_e={}&-66+S_e+3(n_{H_u}+n_{H_d})
+8(\xi_u+\xi_c+\xi_t)\nonumber\\
&+2(\xi_d+\xi_s+\xi_b)
+6(\xi_e+\xi_\mu+\xi_\tau).
\label{app:befinal}
\end{align}
Finally,
\begin{equation}
d_e\equiv(\partial_s\ln\alpha_{\rm EM})_0
 =-\frac{\alpha_{{\rm EM},0}}{3\pi\sqrt6}\,B_e,
 \label{app:definal}
\end{equation}
so $d_e=0$ is precisely the vanishing of the bracket displayed in
Eq.~\eqref{eq:decondition}.

The coefficient $-63/2$ assumes $p_W=1$. More generally,
$p_W=x_{v,0}\equiv(d\ln v/d\ln m_{3/2})_0$ at this order, and the constant
in Eq.~\eqref{app:befinal} becomes
\begin{equation}
 -66\ \longrightarrow\
 -66-\frac{63}{2}(x_{v,0}-1),
 \label{app:vevshift}
\end{equation}
provided the remaining superpartner and heavy-Higgs slopes stay equal to
one. Additional VEV dependence of their eigenmasses must instead be
included directly through Eq.~\eqref{app:thresholdrule}.

\subsection{QCD charge}

Evaluating equation~\eqref{app:uvmatching} for the QCD coupling gives the high-energy color bracket
$S_g-9$ where 
\begin{equation}
 S_g\equiv\sum_{i=1}^3
 (2n_{Q_i}+n_{U_i}+n_{D_i}).
 \label{app:sgdef}
\end{equation}
The twelve complex squark triplets contribute $2$ to $b_3$,
and the adjoint gluino contributes another $2$. Integrating them out with
unit slope gives
\begin{equation}
 \Delta b_3^{\rm SUSY}=-4,
 \qquad
 (S_g-9)-3\Delta b_3^{\rm SUSY}=S_g+3.
 \label{app:qcdsusy}
\end{equation}
Each heavy Dirac quark has $\Delta b_3=-2/3$ and therefore adds
$2\xi_f$ to this bracket. Integrating out $c,b,t$, while retaining
$u,d,s$ in the three-flavor theory, yields
\begin{equation}
 (\partial_s g_3^{-2})_0
 =\frac{\sqrt{2/3}}{16\pi^2}
 \left[S_g+3+2(\xi_c+\xi_b+\xi_t)\right].
 \label{app:g3final}
\end{equation}
Using
$g_3^{-2}=9\ln(\mu/\Lambda_3)/(8\pi^2)$ at fixed $\mu$ gives
\begin{equation}
 d_g\equiv(\partial_s\ln\Lambda_3)_0
 =-\frac{1}{9\sqrt6}(3+S_g+2(\xi_c+\xi_b+\xi_t)),
 \label{app:dgfinal}
\end{equation}
which is Eq.~\eqref{eqn:dg}. Here $\Lambda_3$ is the three-flavor QCD scale and the
strange quark has not been integrated out.
 
This calculation uses one-loop running and assumes that the
tree-level gauge kinetic function and reference cutoff do not
depend on the modulus. The specified superpartner and heavy-Higgs
masses are taken to scale locally as $m_{3/2}$.
The light-quark terms in Eq.~\eqref{app:befinal} are approximate.
Below confinement, a precise calculation must include how hadrons
affect the electromagnetic coupling. Higher-loop and spectrum-dependent threshold corrections
shift both charge relations and require the full mass spectrum.

\section{Vacuum stability in a singlet extension}
\label{app:nmssm_ccb}

The singlet extension of the Higgs sector offers a possible
way to avoid the deeper charge- and color-breaking minima
present in the MSSM solution.
A restricted generalized NMSSM with $\mu=\kappa=0$ is defined
by the superpotential
\begin{equation}
W=W_{\rm MSSM}\big|_{\mu=0}
+\lambda S H_uH_d+\frac12 M_S S^2+\xi_F S,
\end{equation}
with canonical singlet K\"ahler potential $K_S=S\bar S$.
These interactions preserve supersymmetry, which is broken
predominantly by the no-scale modulus. The corresponding
singlet soft terms are
\begin{equation}
\begin{aligned}
V_{\rm soft}^{S}={}&m_S^2|S|^2+
\left(\lambda A_\lambda S H_uH_d
+\frac12 b_S S^2+t_S S+\mathrm{h.c.}\right),
\end{aligned}
\end{equation}
where $b_S\equiv B_S M_S$. The quadratic and linear superpotential coefficients scale as
$\mathcal M_S(T)\propto T^{a_S}$ and
$\mathcal X_F(T)\propto T^{a_\xi}$, independently of the
canonical singlet metric. The fit gives
$a_S\simeq0.3506$ and $a_\xi\simeq-1.0044$, implying
$M_S/m_{3/2}\propto m_{3/2}^{-0.2337}$ and
$\xi_F/m_{3/2}^2\propto m_{3/2}^{-0.3304}$.

For the zero-temperature analysis, an illustrative solution
with $m_{3/2,0}=100~{\rm TeV}$, $\tan\beta_0=2$, and
$\lambda\simeq0.5937$ has the present matching-scale parameters
\begin{equation}
\frac{M_S}{m_{3/2}}=-0.55,\quad
\frac{m_S^2}{m_{3/2}^2}=1,\quad
\frac{b_S}{m_{3/2}^2}\simeq-0.71.
\label{eq:nmssm_benchmark}
\end{equation}
together with
\begin{equation}
\begin{aligned}
\frac{A_\lambda}{m_{3/2}}\simeq1.524,\quad
\frac{\xi_F}{m_{3/2}^2}\simeq-0.241,
\quad
\frac{t_S}{m_{3/2}^3}\simeq-0.725.
\end{aligned}
\end{equation}
The holomorphic singlet coupling obeys
$\lambda(T)\propto T^{\kappa_\lambda}$,
where $\kappa_\lambda\simeq-0.07457$ is fixed by
$A_\lambda$ and the modular weights.
The Higgs-sector soft parameters are
\begin{equation}
\frac{m_{H_u}^2}{m_{3/2}^2}=-\frac18,
\qquad
\frac{m_{H_d}^2}{m_{3/2}^2}=\frac12,
\qquad b=0.
\end{equation}
The matter-field modular weights for the three generations $i=1,2,3$ are
\begin{equation}
n_{Q_i}=n_{L_i}=\tfrac12,\quad
n_{D_i}=-3,\quad
n_{U_i}=n_{E_i}=\tfrac34.
\label{b6}
\end{equation}
while $n_{H_u}=9/8$ and $n_{H_d}=1/2$.  Consequently, all diagonal matter scalar soft masses
$\widetilde m_I^2=(1-n_I)m_{3/2}^2$ are positive, with the smallest entries
$m_{U_i}^2=m_{E_i}^2=m_{3/2}^2/4$. The holomorphic Yukawa exponents were adjusted simultaneously
so that the light-fermion and gauge-sector scalar charges
are numerically consistent with zero in the one-loop
matching calculation.

\begin{figure}[ht]
\centering
\hspace*{-5mm}
\includegraphics[width=\columnwidth,
height=0.65\columnwidth
]{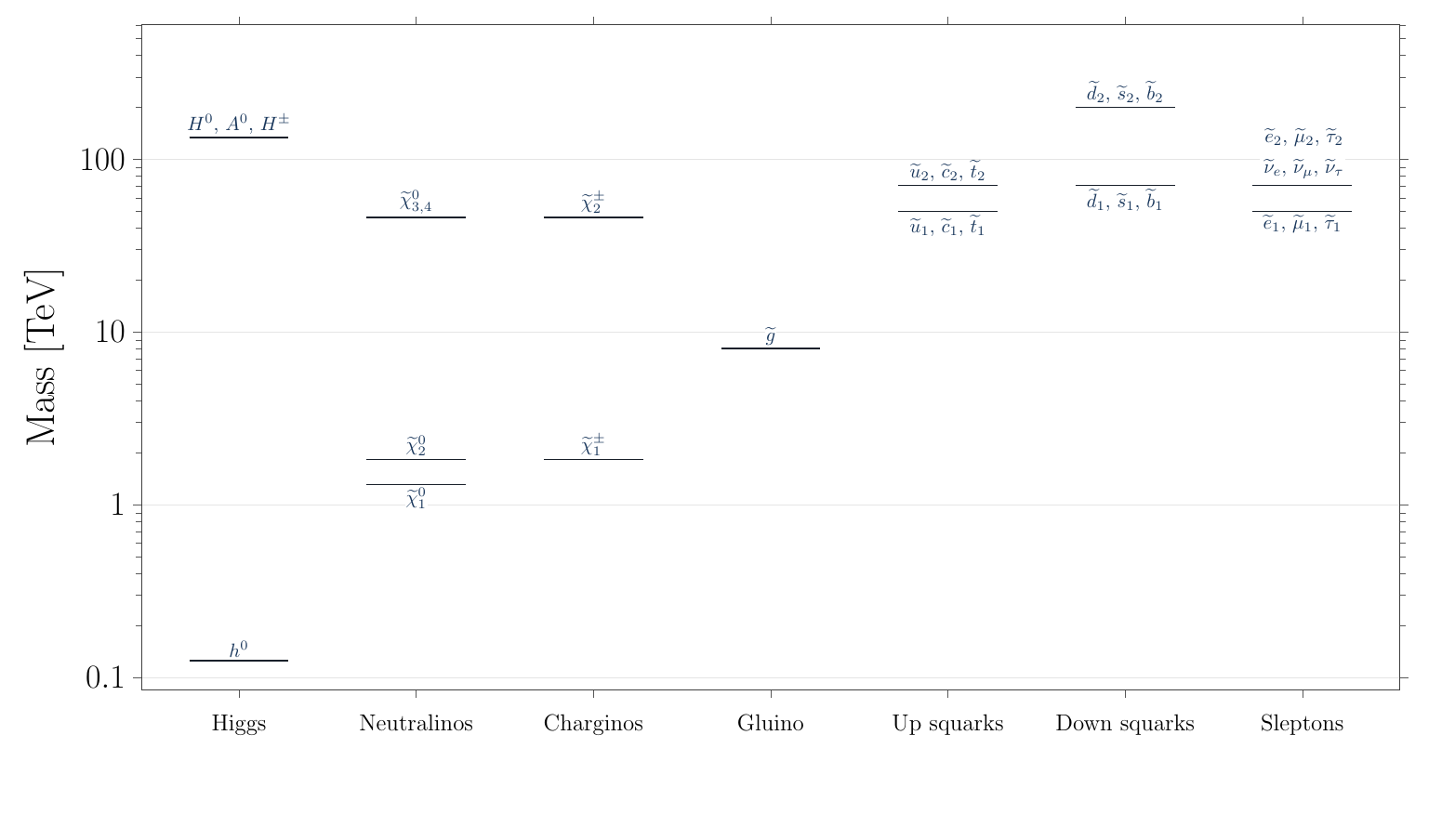}
\caption{Present-day MSSM spectrum for the parameters of
Fig.~\ref{vm32evol} and Appendix~\ref{app:finiteT},
evaluated at $Q=m_{3/2}$, except for the light Higgs
shown at its approximate pole mass.
}
\vspace*{-1.5mm}
\label{mssmspec}
\end{figure}
\begin{figure}[ht]
\centering
\hspace*{-5mm}
\includegraphics[width=\columnwidth,height=0.65\columnwidth]{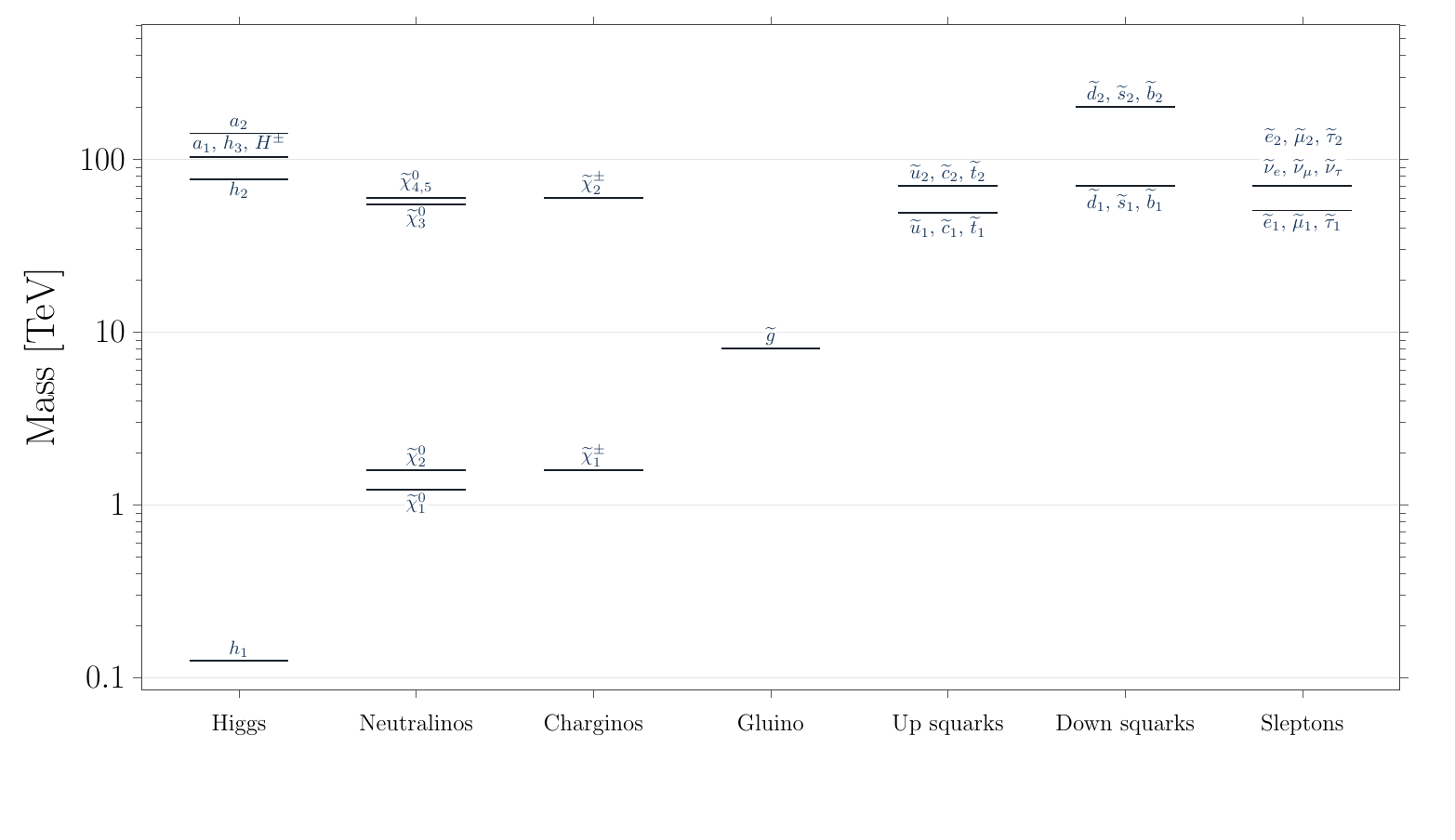}
\caption{Present-day NMSSM spectrum for
Eqs.~\eqref{eq:nmssm_benchmark}--\eqref{b6}, with Yukawa
exponents from Appendix~\ref{app:finiteT}, evaluated at
$Q=m_{3/2}$, except for the light Higgs shown at its
approximate pole mass. 
}
\vspace*{-1.5mm}
\label{nmssmspec}
\end{figure}

The singlet interaction supplies the additional Higgs quartic
$\lambda^2|H_uH_d|^2$. Together with the singlet mixing and
radiative corrections, this permits electroweak symmetry
breaking and an approximately $125~{\rm GeV}$ Higgs while retaining
positive sfermion soft masses and moderate trilinear mixing.
The present singlet expectation value,
$\langle S\rangle_0\simeq100.88~{\rm TeV}$,
is of order the supersymmetry-breaking scale, and the two
predominantly singlet scalar squared masses are positive. The Higgs mass was obtained using approximate one-loop
supersymmetric matching, SM running, and the SMDR~\cite{Martin:2019lqd} pole-mass
conversion at two loops with selected leading three-loop
contributions. This gives an approximate $125~{\rm GeV}$
solution, rather than a full two-loop NMSSM pole-mass calculation.

\begin{figure}[ht]
\centering
\hspace*{-5mm}
\includegraphics[width=\columnwidth]{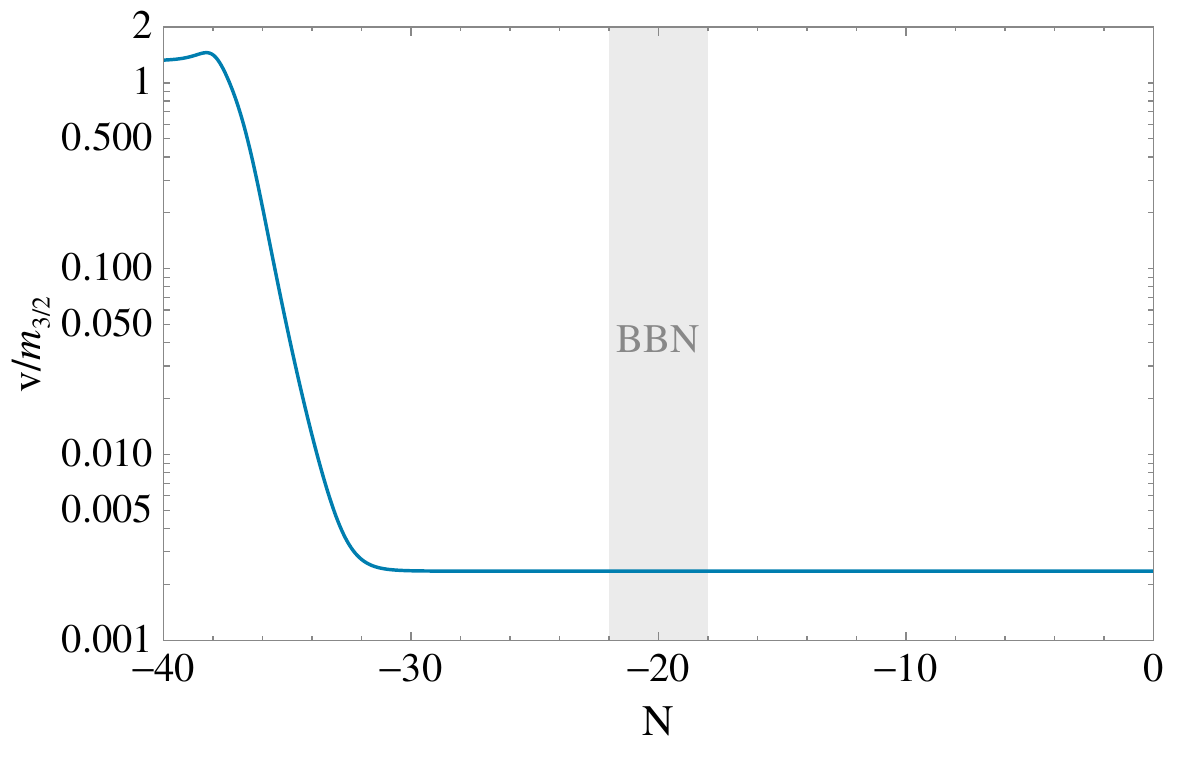}
\caption{The evolution of $v/m_{3/2}$ with e-folds $N$ for the NMSSM solution,
using the stationary Higgs VEV at $Q=m_{3/2}$.
}
\vspace*{-1.5mm}
\label{nmssmv}
\end{figure}

Vacuum stability was tested at selected values in the range
$-40\leq N\leq0$, allowing all real and imaginary scalar
components to vary, including all generations and colors.
No configuration below the neutral electroweak minimum was
found, and all sfermion and physical charged-Higgs squared
masses were positive at the tested epochs.
These finite searches do not constitute a proof of global
stability. Figure~\ref{nmssmv} shows the corresponding evolution
of $v/m_{3/2}$ at $Q=m_{3/2}$,
which reproduces $v_0\simeq246~{\rm GeV}$ after matching
to the weak scale.

These results illustrate that the leading scalar-charge
cancellation conditions can coexist with a neutral electroweak
minimum and positive matter-field soft masses in a singlet
extension, with no lower CCB configuration found in the
searches performed. The qualitative evolution of
$v/m_{3/2}$ remains similar to that in
Sec.~\ref{sec:modelconstraints}.

\section{Effects of finite temperature corrections}
\label{app:finiteT}

Finite-temperature corrections to Eq.~\eqref{eq:simplepot}
modify the evolution of $\sigma$.  They are negligible today at the calculated photon temperature $T_{\gamma 0}\simeq 2.8\,{\rm K}$, but can be important when the temperature exceeds the particle masses. The $\sigma$-dependent
masses generate terms such as $m^2(\sigma)T^2$ in the
high-temperature expansion. The temperature follows comoving
entropy conservation, $s(T,n,\sigma)a^3=\mathrm{constant}$,
where $s$ is the entropy density, including the conserved
asymmetry and changes in particle abundances. After neutrino
decoupling, entropy is conserved separately in the coupled
plasma and neutrino sectors.

The particle masses are evaluated at the approximate
temperature-dependent minimum of the neutral scalar fields,
with charged and colored scalar VEVs set to zero.
The resulting thermal free energy modifies the evolution of $\sigma$ and hence the Higgs VEV.
At one loop, it is given by
\begin{equation}
\Delta F_{\rm th}^{(1)}(\sigma,T)
=\frac{T^4}{2\pi^2}
\sum_i(-1)^{F_i}g_iJ_i(y_i),
\label{eq:thermalfree}
\end{equation}
where $g_i$ counts the internal degrees of freedom,
$F_i=0$ ($1$) for bosons (fermions),
$y_i\equiv m_i^2(\sigma, T)/T^2$, and
$J_{B,F}(y)\equiv\int_0^\infty dx\,x^2
\ln\left(1\mp e^{-\sqrt{x^2+y}}\right)$,
with $J_i=J_B$ ($J_F$) for bosons (fermions).
The numerical evolution also includes selected two-loop
interaction terms of order $g^2T^4$ and $y^2T^4$,
finite-mass corrections, and ring contributions accounting
for thermal screening. Confinement is modeled by smoothly suppressing quark and
gluon thermal contributions near $155~{\rm MeV}$.
The total $\Delta F_{\rm th}$ includes these terms and
the change in Higgs-sector vacuum energy between the
approximate thermal and zero-temperature minima.
A complete two-loop treatment is beyond this analysis. In the early MSSM plasma, the two heavy Higgs supermultiplets
are integrated out with finite one-loop thresholds.
The remaining parameters are run to
$Q_{\rm th}=(T^4+m_{3/2}^4)^{1/4}$, and the resulting
free energy is smoothly joined to the massive calculation.

Using the thermal free energy
$\Delta F_{\rm th}$, the pressure and energy density are
$P_{\rm th}=-\Delta F_{\rm th}$ and
$\rho_{\rm th}=\Delta F_{\rm th}
-T(\partial_T\Delta F_{\rm th})_\sigma$.
The thermal energy density enters the Friedmann constraint,
while $4\Omega_{r0}e^{-4N}$ in \eqref{eq:eom} is replaced by
$(\rho_{\rm th}+P_{\rm th})/(M_P^2H_0^2)$.
Neutrino energy and pressure are included separately after
decoupling. The force
$-(\partial_\sigma\Delta F_{\rm th})_T$
sources the scalar evolution.

The conserved matter asymmetry adds the excess free-energy density
$\Delta{\cal F}_b(T,n,\sigma)\equiv
{\cal F}(T,n,\sigma)-{\cal F}(T,0,\sigma)$
with conserved number density $n\propto e^{-3N}$, giving
$P_b=n\partial_n\Delta{\cal F}_b-\Delta{\cal F}_b$,
$\rho_b=\Delta{\cal F}_b-T\partial_T\Delta{\cal F}_b$,
and $F_b=-\partial_\sigma\Delta{\cal F}_b$,
holding other arguments fixed.
Quark and hadron descriptions are smoothly interpolated near confinement.
For nonrelativistic baryons,
$\Delta{\cal F}_b\simeq n\,m_B(\sigma)$,
so
$\rho_b\simeq\rho_{b0}e^{-3N}
m_B(\sigma)/m_B(\sigma_0)$
and $F_b\simeq-\rho_b\partial_\sigma\ln m_B$, where the mass per baryon $m_B$ is estimated from the calculated QCD scale,
light-quark and electron masses, and electromagnetic coupling,
assuming a fixed hydrogen--helium mixture.
The asymmetry contributes $\rho_b$ to the Friedmann constraint
and $(\rho_b+P_b)/(M_P^2H_0^2)$ to the second equation
in Eq.~\eqref{eq:eom}. To avoid double counting,
$\Omega_{m0}$ is replaced by $\Omega_{{\rm dm},0}=0.2659$,
with $\Omega_{b0}=0.049$.
The nonthermal examples include the asymmetry source
but omit the equilibrium thermal contribution.

For the MSSM thermal evolution, the Higgs-sector inputs are held fixed:
\begin{equation}
\frac{\mu_0}{m_{3/2,0}}\simeq 0.16,\quad
C_{ud,0}\simeq 0.07,\quad
n_\mu\simeq -3.62.
\label{eq:mssm_thermal_inputs}
\end{equation}
The cosmological initial conditions are adjusted, with the
scalar-potential normalization held fixed, to obtain
\begin{equation}
\sigma(0)=-9M_P,\qquad
\sigma'(0)\simeq -3.8\times10^{-12}M_P.
    \label{bcmssm}
\end{equation} 
The matter-field modular weights are given in Eq.~\eqref{b6},
while $n_{H_u}=9/8$ and $n_{H_d}=-1/2$,
giving positive tree-level sfermion soft masses.
The numerical Yukawa exponents are chosen to suppress
the present-day light-fermion and gauge-sector scalar
charges, including one-loop threshold corrections and
SM running. Their values are listed below, rounded to
four decimal places.
\begin{center}
\begin{tabular}{c|rr}
\hline
$f$ & $\kappa_f$ (MSSM) & $\kappa_f$ (NMSSM) \\
\hline
$u$    &  -2.7082 &  0.0013 \\
$c$    & -14.7541 & -0.2875 \\
$t$    &   0.3125 & -0.1902 \\
$d$    &  -6.5548 &  2.4769 \\
$s$    &  -6.5531 &  2.4769 \\
$b$    &   3.0000 &  2.4666 \\
$e$    &  -8.2929 &  0.5573 \\
$\mu$  &  -2.5023 &  0.1750 \\
$\tau$ &   0.7500 & -0.0039 \\
\hline
\end{tabular}
\end{center}
The same Yukawa exponents apply to the corresponding
zero-temperature solutions, with full-precision inputs
used throughout. The charges are consistent with zero within
a numerical resolution of a few $10^{-5}$ using one-loop
supersymmetric matching and SM running. A more accurate
determination of the scalar-charge cancellation requires
improved numerical precision and control of higher-order
matching corrections, including the Higgs quadratic
coefficient and its modulus dependence.
  
\begin{figure}[ht]
\centering
\hspace*{-5mm}
\includegraphics[width=\columnwidth,height=0.65\columnwidth]{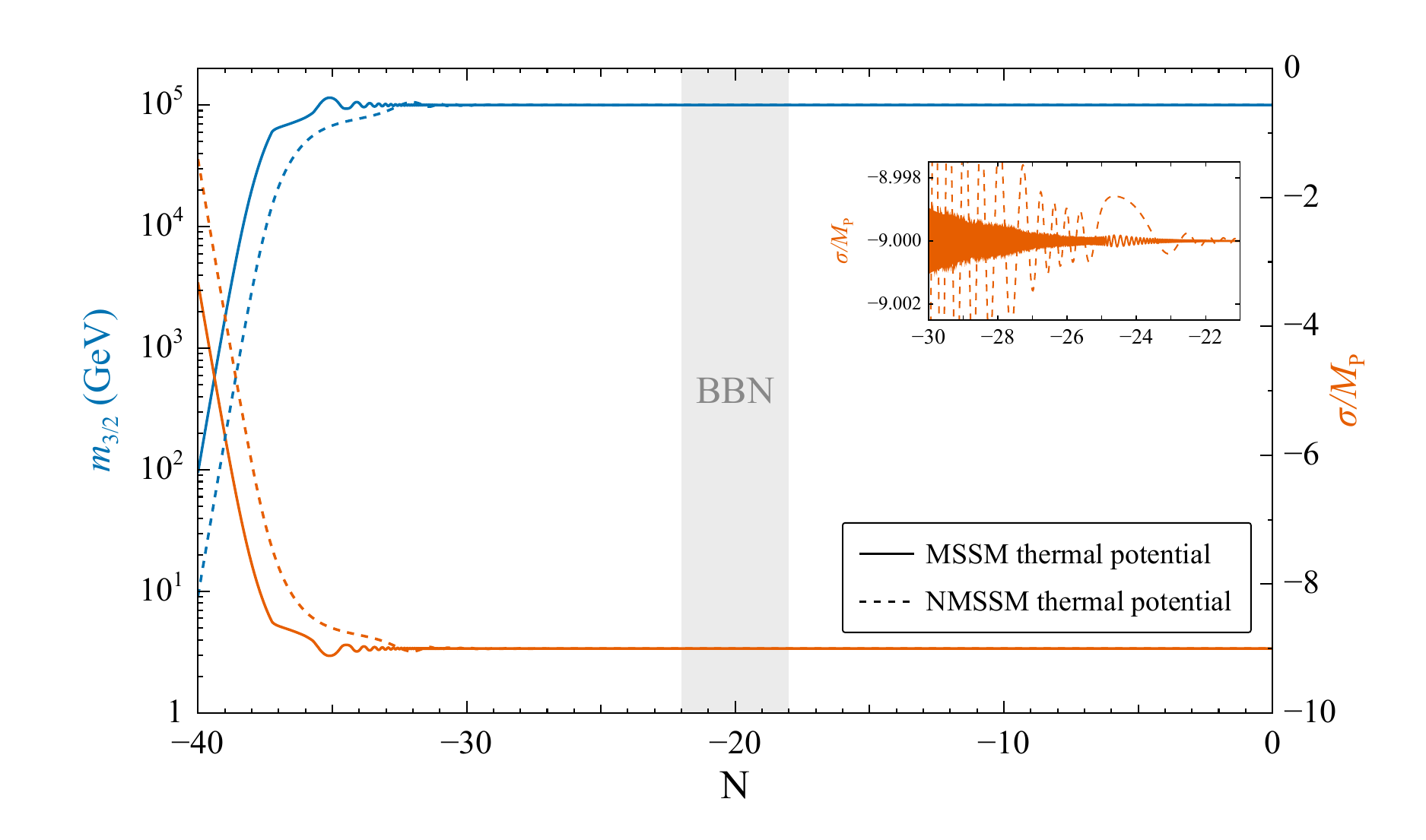}
\caption{The evolution of $m_{3/2}$ (blue) and $\sigma/M_P$
(orange), including thermal and baryonic forces, for
the MSSM (solid) and NMSSM (dashed), with boundary conditions
\eqref{bcmssm} and \eqref{eq:bcnmssm}, respectively.}
\vspace*{-1.5mm}
\label{m32sigth}
\end{figure}

\begin{figure}[ht]
\centering
\hspace*{-5mm}
\includegraphics[width=\columnwidth]{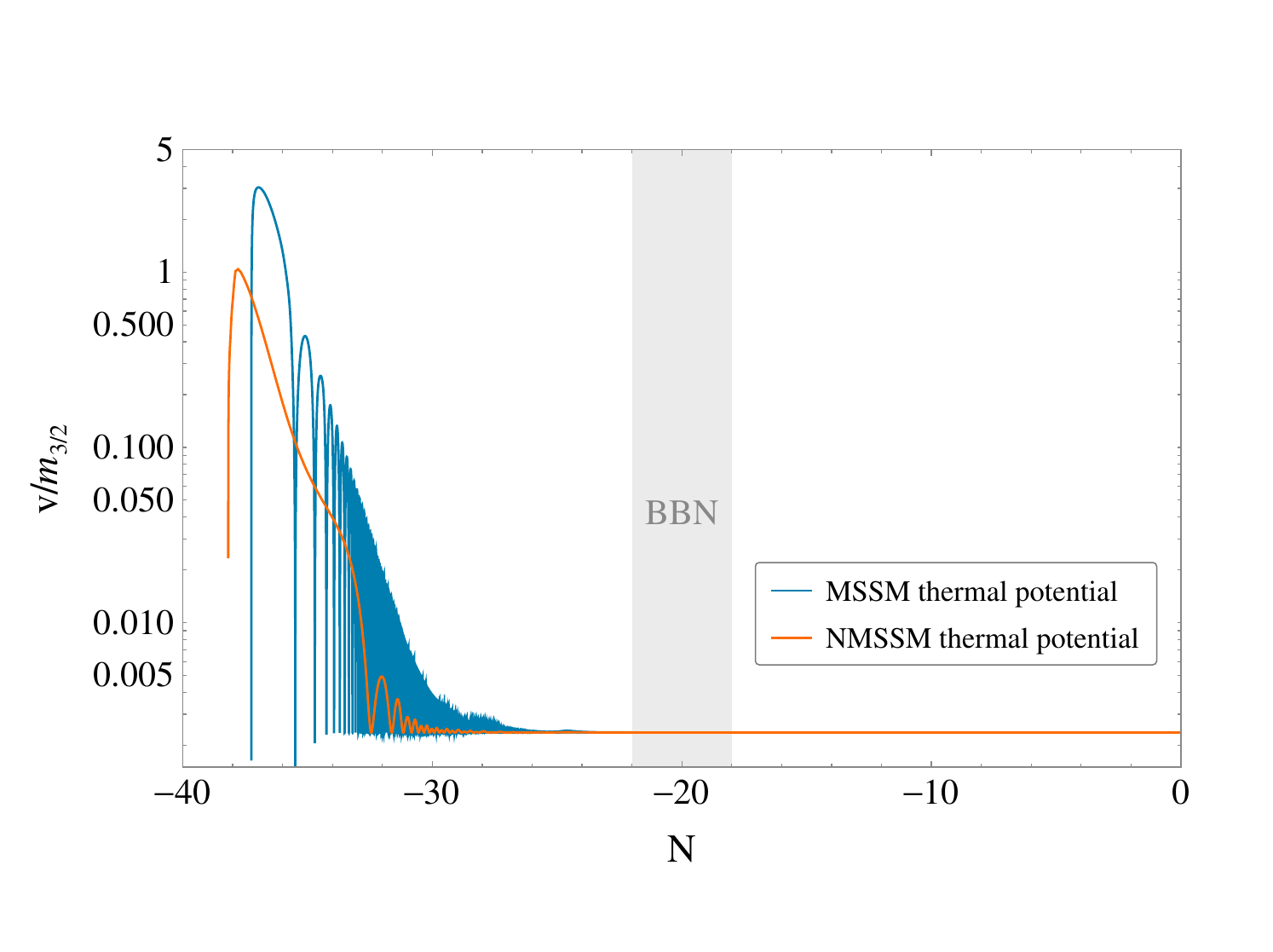}
\caption{The evolution of $v/m_{3/2}$ with e-fold number $N$ for the thermal MSSM and
NMSSM solutions, evaluated at the approximate interpolated thermal Higgs
minimum at $Q=m_{3/2}$. Intervals with $v=0$ are omitted.}
\vspace*{-1.5mm}
\label{vm32th}
\end{figure}

Figure~\ref{m32sigth} shows the evolution of $m_{3/2}$ and
$\sigma$ including thermal and baryonic forces. Early rolling
is followed by damped oscillations and a nearly stationary
present endpoint. The equilibrium thermal force becomes
negligible at late times, while the baryonic force remains.
At $N=-40$, the MSSM solution has
$m_{3/2}\simeq95~{\rm GeV}$ and unbroken electroweak symmetry.
The ratio $v/m_{3/2}$ subsequently reaches approximately $3$
before relaxing toward its present value, as shown in
Fig.~\ref{vm32th}, while $m_{3/2}$ approaches $100~{\rm TeV}$
today. During BBN ($-22\leq N\leq-18$), the maximum fractional
departures of $m_{3/2}$ and the running VEV at $Q=m_{3/2}$
from their present values are approximately
$2.1\times10^{-5}$ and $2.4\times10^{-4}$, respectively.


\begin{figure}[tbp]
\centering
\hspace*{-5mm}
\includegraphics[width=\columnwidth,height=0.8\columnwidth]{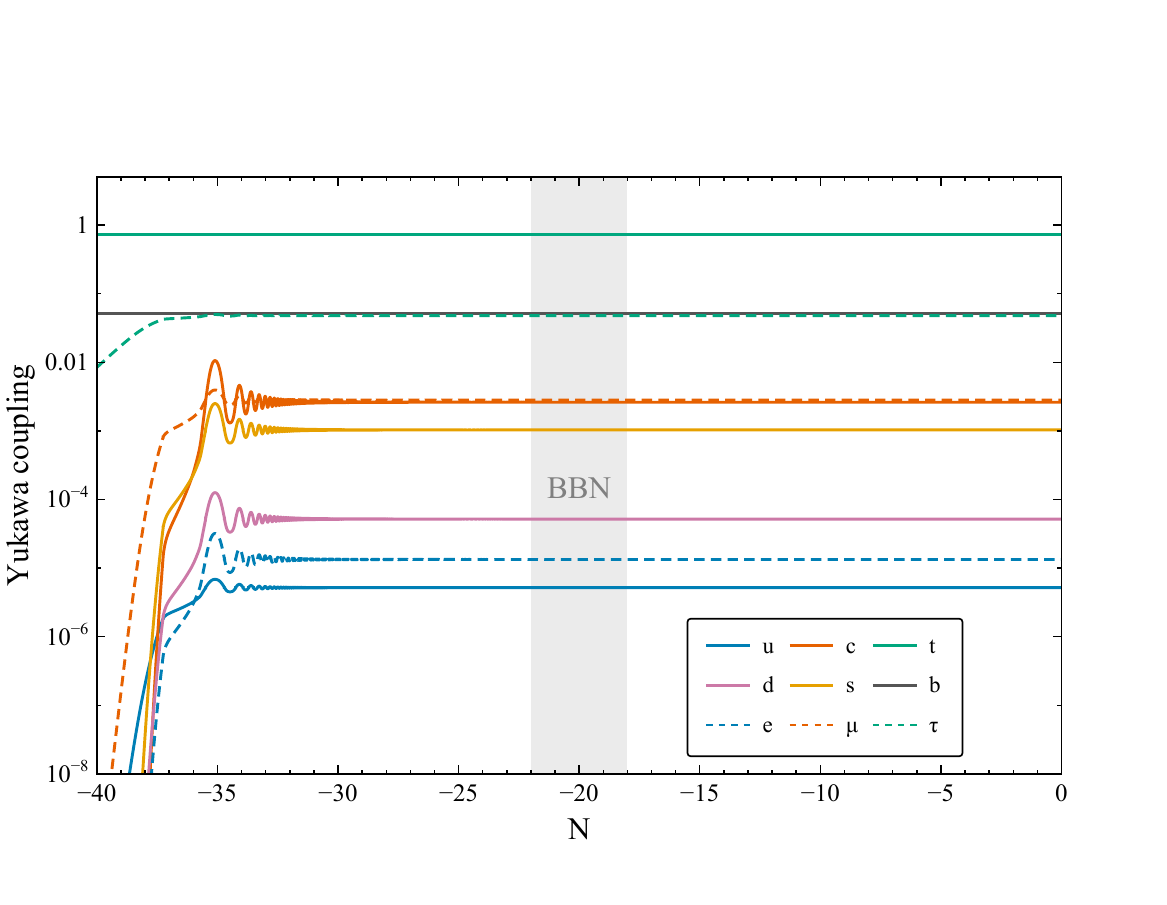}
\caption{The evolution of the canonically normalized Yukawa couplings
at $Q=m_{3/2}(\sigma(N))$ for the MSSM solution along the thermal trajectory.
}
\vspace*{-1.5mm}
\label{yukth}
\end{figure} 
Our parameter choices yield an approximate matched Higgs pole mass of
$125$~GeV. As in the nonthermal case, deeper CCB configurations
occur in the zero-temperature potential. 
The solution starts in an   electroweak-symmetric
phase at $N=-40$ and encounters intervals of broken
electroweak symmetry during its subsequent oscillations.
The present-day tree-level bounce estimates suggest suppressed
zero-temperature tunnelling along the examined charge- or
color-breaking directions. Thermal tunnelling and vacuum
survival over the cosmological history have not been calculated.

As shown in Fig.~\ref{m32sigth}, early rolling of $\sigma$
is followed by damped oscillations and a slowly varying
regime near the present endpoint. The evolution is shown
over $-40\leq N\leq0$. Extending the solution to earlier times would require a separate analysis of the Higgs branch and its stability. The Yukawa couplings in Fig.~\ref{yukth} remain perturbative throughout the displayed interval.

\begin{figure}[tbp]
\centering
\hspace*{-5mm}
\includegraphics[width=\columnwidth,height=0.8\columnwidth]{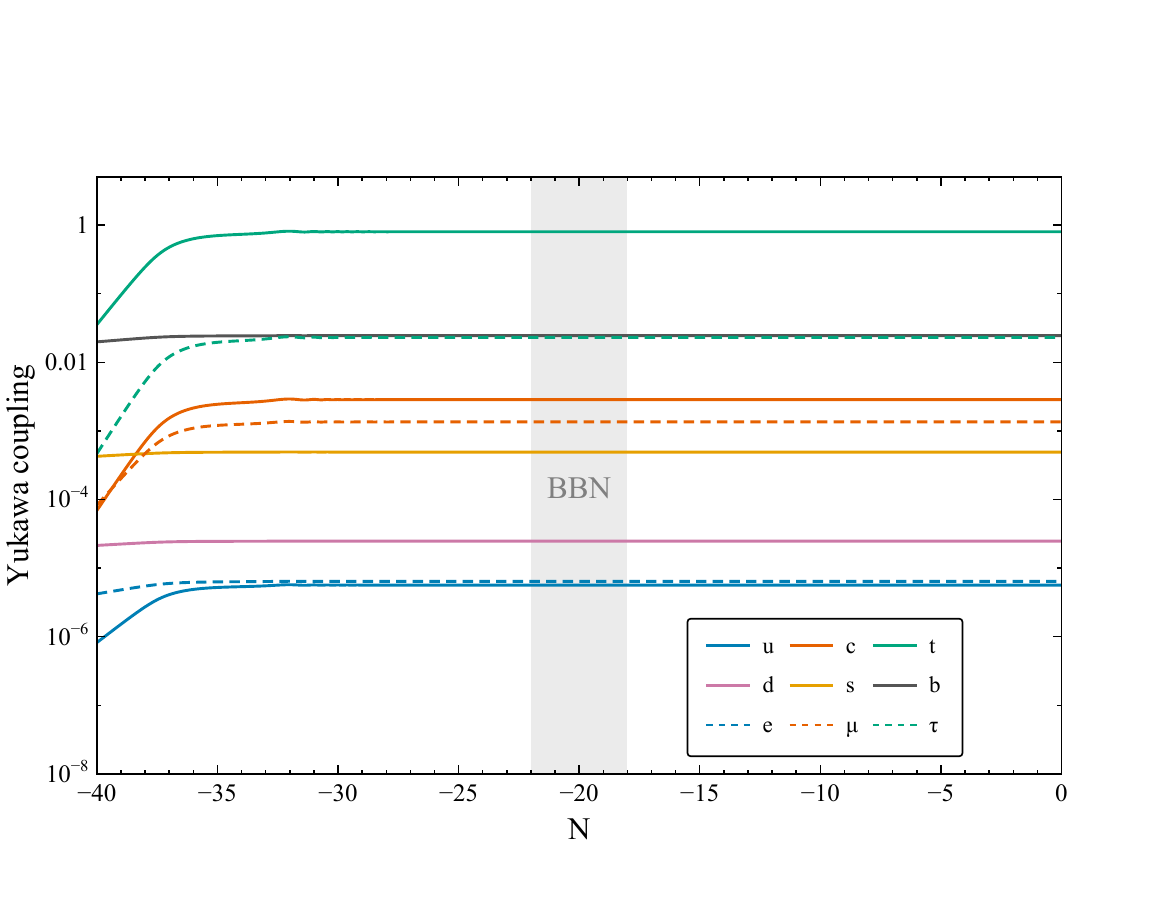}
\caption{The evolution of the canonically normalized Yukawa couplings
at $Q=m_{3/2}(\sigma(N))$ for the NMSSM solution along the thermal trajectory.}
\vspace*{-1.5mm}
\label{yukthnmssm}
\end{figure}

The previous MSSM solution admits deeper CCB minima.
Since no such minimum was found in the zero-temperature
searches for the singlet solution of App.~\ref{app:nmssm_ccb},
we also consider its thermal evolution with the parameters
in Eqs.~\eqref{eq:nmssm_benchmark}--\eqref{b6} unchanged.
The scalar boundary conditions are
\begin{equation}
\sigma(0)=-9M_P,\qquad
\sigma'(0)\simeq-1.96\times10^{-12}M_P.
\label{eq:bcnmssm}
\end{equation}
The small present-day velocities require tuning that depends
on the approximate treatment of baryons and QCD.
The Higgs pole-mass estimate remains approximately
$125~\mathrm{GeV}$ today. Figure~\ref{vm32th} shows the
corresponding evolution of $v/m_{3/2}$, including early
oscillations induced by thermal effects. The Yukawa
couplings remain perturbative over $-40\leq N\leq0$,
as shown in Fig.~\ref{yukthnmssm}.

In summary, the zero-temperature quintessence potential has no finite
minimum. Thermal effects produce damped oscillations and
become negligible as the plasma cools, while the baryonic
force continues to influence the field.
The quintessence potential supplies the dark-energy density even when
baryons dominate the scalar force.

\bibliographystyle{JHEP}
\bibliography{ref}

@article{Brax:2006dc,
    author = "Brax, Philippe and Martin, Jerome",
    title = "{Dark Energy and the MSSM}",
    eprint = "hep-th/0605228",
    archivePrefix = "arXiv",
    doi = "10.1103/PhysRevD.75.083507",
    journal = "Phys. Rev. D",
    volume = "75",
    pages = "083507",
    year = "2007"
}

@article{Brignole:1997wnc,
    author = "Brignole, A. and Ibanez, Luis E. and Munoz, C.",
    title = "{Soft supersymmetry breaking terms from supergravity and superstring models}",
    eprint = "hep-ph/9707209",
    archivePrefix = "arXiv",
    reportNumber = "CERN-TH-97-143, FTUAM-97-7, KAIST-TH-97-7",
    doi = "10.1142/9789812839657_0003",
    journal = "Adv. Ser. Direct. High Energy Phys.",
    volume = "18",
    pages = "125--148",
    year = "1998"
}

@article{Bertotti:2003rm,
    author = "Bertotti, B. and Iess, L. and Tortora, P.",
    title = "{A test of general relativity using radio links with the Cassini spacecraft}",
    doi = "10.1038/nature01997",
    journal = "Nature",
    volume = "425",
    pages = "374--376",
    year = "2003"
}

@article{Brax:2006np,
    author = "Brax, Philippe and Martin, Jerome",
    title = "{Moduli Fields as Quintessence and the Chameleon}",
    eprint = "hep-th/0612208",
    archivePrefix = "arXiv",
    doi = "10.1016/j.physletb.2007.02.019",
    journal = "Phys. Lett. B",
    volume = "647",
    pages = "320--329",
    year = "2007"
}

@article{Brax:1999gp,
    author = "Brax, Philippe and Martin, Jerome",
    title = "{Quintessence and supergravity}",
    eprint = "astro-ph/9905040",
    archivePrefix = "arXiv",
    doi = "10.1016/S0370-2693(99)01209-5",
    journal = "Phys. Lett. B",
    volume = "468",
    pages = "40--45",
    year = "1999"
}

@article{Chiang:2018jdg,
    author = "Chiang, Chien-I and Murayama, Hitoshi",
    title = "{Building Supergravity Quintessence Model}",
    eprint = "1808.02279",
    archivePrefix = "arXiv",
    primaryClass = "hep-th",
    month = "8",
    year = "2018"
}

@article{Cremmer:1983bf,
    author = "Cremmer, E. and Ferrara, S. and Kounnas, C. and Nanopoulos, Dimitri V.",
    title = "{Naturally Vanishing Cosmological Constant in N=1 Supergravity}",
    reportNumber = "CERN-TH-3667",
    doi = "10.1016/0370-2693(83)90106-5",
    journal = "Phys. Lett. B",
    volume = "133",
    pages = "61",
    year = "1983"
}

@article{Ellis:1984bm,
    author = "Ellis, John R. and Kounnas, C. and Nanopoulos, Dimitri V.",
    title = "{No Scale Supersymmetric Guts}",
    reportNumber = "CERN-TH-3824",
    doi = "10.1016/0550-3213(84)90555-8",
    journal = "Nucl. Phys. B",
    volume = "247",
    pages = "373--395",
    year = "1984"
}

@article{Uzan:2002vq,
    author = "Uzan, Jean-Philippe",
    title = "{The Fundamental Constants and Their Variation: Observational Status and Theoretical Motivations}",
    eprint = "hep-ph/0205340",
    archivePrefix = "arXiv",
    doi = "10.1103/RevModPhys.75.403",
    journal = "Rev. Mod. Phys.",
    volume = "75",
    pages = "403",
    year = "2003"
}

@article{Barreiro:1999zs,
    author = "Barreiro, T. and Copeland, Edmund J. and Nunes, N. J.",
    title = "{Quintessence arising from exponential potentials}",
    eprint = "astro-ph/9910214",
    archivePrefix = "arXiv",
    reportNumber = "SUSX-TH-016",
    doi = "10.1103/PhysRevD.61.127301",
    journal = "Phys. Rev. D",
    volume = "61",
    pages = "127301",
    year = "2000"
}

@article{Rubano:2001su,
    author = "Rubano, Claudio and Scudellaro, Paolo",
    title = "{On some exponential potentials for a cosmological scalar field as quintessence}",
    eprint = "astro-ph/0103335",
    archivePrefix = "arXiv",
    doi = "10.1023/A:1015395512123",
    journal = "Gen. Rel. Grav.",
    volume = "34",
    pages = "307--328",
    year = "2002"
}

@article{Sen:2001xu,
    author = "Sen, A. A. and Sethi, S.",
    title = "{Quintessence model with double exponential potential}",
    eprint = "gr-qc/0111082",
    archivePrefix = "arXiv",
    doi = "10.1016/S0370-2693(02)01547-2",
    journal = "Phys. Lett. B",
    volume = "532",
    pages = "159--165",
    year = "2002"
}

@article{Ibanez:1992hc,
    author = "Ibanez, Luis E. and Lust, Dieter",
    title = "{Duality anomaly cancellation, minimal string unification and the effective low-energy Lagrangian of 4-D strings}",
    eprint = "hep-th/9202046",
    archivePrefix = "arXiv",
    reportNumber = "CERN-TH-6380-92",
    doi = "10.1016/0550-3213(92)90189-I",
    journal = "Nucl. Phys. B",
    volume = "382",
    pages = "305--361",
    year = "1992"
}

@article{Eroncel:2025bcb,
    author = {Er{\"o}ncel, Cem and Gouttenoire, Yann and Sato, Ryosuke and Servant, G{\'e}raldine and Simakachorn, Peera},
    title = "{Universal Bound on the Duration of a Kination Era}",
    eprint = "2501.17226",
    archivePrefix = "arXiv",
    primaryClass = "hep-ph",
    doi = "10.1103/k7ty-gwjg",
    journal = "Phys. Rev. Lett.",
    volume = "135",
    number = "10",
    pages = "101002",
    year = "2025"
}

@article{Visinelli:2009kt,
    author = "Visinelli, Luca and Gondolo, Paolo",
    title = "{Axion cold dark matter in non-standard cosmologies}",
    eprint = "0912.0015",
    archivePrefix = "arXiv",
    primaryClass = "astro-ph.CO",
    doi = "10.1103/PhysRevD.81.063508",
    journal = "Phys. Rev. D",
    volume = "81",
    pages = "063508",
    year = "2010"
}

@article{Pisarski:1983ms,
    author = "Pisarski, Robert D. and Wilczek, Frank",
    title = "{Remarks on the Chiral Phase Transition in Chromodynamics}",
    reportNumber = "NSF-ITP-83-152",
    doi = "10.1103/PhysRevD.29.338",
    journal = "Phys. Rev. D",
    volume = "29",
    pages = "338--341",
    year = "1984"
}

@article{Ellis:2013nxa,
    author = "Ellis, John and Nanopoulos, Dimitri V. and Olive, Keith A.",
    title = "{Starobinsky-like Inflationary Models as Avatars of No-Scale Supergravity}",
    eprint = "1307.3537",
    archivePrefix = "arXiv",
    primaryClass = "hep-th",
    reportNumber = "KCL-PH-TH-2013-13, LCTS-2013-22, CERN-PH-TH-2013-156, ACT-6-13, UMN-TH-3213-13, FTPI-MINN-13-23, MIFPA-13-22",
    doi = "10.1088/1475-7516/2013/10/009",
    journal = "JCAP",
    volume = "10",
    pages = "009",
    year = "2013"
}

@article{Yoo:2002vw,
    author = "Yoo, Jaiyul and Scherrer, Robert J.",
    title = "{Big bang nucleosynthesis and cosmic microwave background constraints on the time variation of the Higgs vacuum expectation value}",
    eprint = "astro-ph/0211545",
    archivePrefix = "arXiv",
    doi = "10.1103/PhysRevD.67.043517",
    journal = "Phys. Rev. D",
    volume = "67",
    pages = "043517",
    year = "2003"
}

@article{Damour:2010rp,
    author = "Damour, Thibault and Donoghue, John F.",
    title = "{Equivalence Principle Violations and Couplings of a Light Dilaton}",
    eprint = "1007.2792",
    archivePrefix = "arXiv",
    primaryClass = "gr-qc",
    doi = "10.1103/PhysRevD.82.084033",
    journal = "Phys. Rev. D",
    volume = "82",
    pages = "084033",
    year = "2010"
}

@article{Bagger:1999rd,
    author = "Bagger, Jonathan A. and Moroi, Takeo and Poppitz, Erich",
    title = "{Anomaly mediation in supergravity theories}",
    eprint = "hep-th/9911029",
    archivePrefix = "arXiv",
    reportNumber = "IASSNS-HEP-99-89, YCTP-P27-99",
    doi = "10.1088/1126-6708/2000/04/009",
    journal = "JHEP",
    volume = "04",
    pages = "009",
    year = "2000"
}

@article{MICROSCOPE:2022doy,
    author = "Touboul, Pierre and others",
    collaboration = "MICROSCOPE",
    title = "{MICROSCOPE Mission: Final Results of the Test of the Equivalence Principle}",
    eprint = "2209.15487",
    archivePrefix = "arXiv",
    primaryClass = "gr-qc",
    doi = "10.1103/PhysRevLett.129.121102",
    journal = "Phys. Rev. Lett.",
    volume = "129",
    number = "12",
    pages = "121102",
    year = "2022"
}

@article{Kaplunovsky:1994fg,
    author = "Kaplunovsky, Vadim and Louis, Jan",
    title = "{Field dependent gauge couplings in locally supersymmetric effective quantum field theories}",
    eprint = "hep-th/9402005",
    archivePrefix = "arXiv",
    reportNumber = "UTTG-94-1, LMU-TPW-94-1",
    doi = "10.1016/0550-3213(94)00150-2",
    journal = "Nucl. Phys. B",
    volume = "422",
    pages = "57--124",
    year = "1994"
}

@article{Brax:1999yv,
    author = "Brax, Philippe and Martin, Jerome",
    title = "{The Robustness of quintessence}",
    eprint = "astro-ph/9912046",
    archivePrefix = "arXiv",
    reportNumber = "SACLAY-SPH-T-99-075",
    doi = "10.1103/PhysRevD.61.103502",
    journal = "Phys. Rev. D",
    volume = "61",
    pages = "103502",
    year = "2000"
}

@article{Giudice:1988yz,
    author = "Giudice, G. F. and Masiero, A.",
    title = "{A Natural Solution to the mu Problem in Supergravity Theories}",
    reportNumber = "SISSA-17/88/EP",
    doi = "10.1016/0370-2693(88)91613-9",
    journal = "Phys. Lett. B",
    volume = "206",
    pages = "480--484",
    year = "1988"
}

@article{Choi:2018rze,
    author = "Choi, Kiwoon and Chway, Dongjin and Shin, Chang Sub",
    title = "{The dS swampland conjecture with the electroweak symmetry and QCD chiral symmetry breaking}",
    eprint = "1809.01475",
    archivePrefix = "arXiv",
    primaryClass = "hep-th",
    doi = "10.1007/JHEP11(2018)142",
    journal = "JHEP",
    volume = "11",
    pages = "142",
    year = "2018"
}

@article{Adelberger:2003zx,
    author = "Adelberger, E. G. and Heckel, Blayne R. and Nelson, A. E.",
    title = "{Tests of the gravitational inverse square law}",
    eprint = "hep-ph/0307284",
    archivePrefix = "arXiv",
    doi = "10.1146/annurev.nucl.53.041002.110503",
    journal = "Ann. Rev. Nucl. Part. Sci.",
    volume = "53",
    pages = "77--121",
    year = "2003"
}

@article{Fields:2019pfx,
    author = "Fields, Brian D. and Olive, Keith A. and Yeh, Tsung-Han and Young, Charles",
    title = "{Big-Bang Nucleosynthesis after Planck}",
    eprint = "1912.01132",
    archivePrefix = "arXiv",
    primaryClass = "astro-ph.CO",
    reportNumber = "UMN--TH--3902/19, FTPI--MINN--19/25",
    doi = "10.1088/1475-7516/2020/03/010",
    journal = "JCAP",
    volume = "03",
    pages = "010",
    year = "2020",
    note = "[Erratum: JCAP 11, E02 (2020)]"
}

@article{Shifman:1986zi,
    author = "Shifman, Mikhail A. and Vainshtein, A. I.",
    title = "{Solution of the Anomaly Puzzle in SUSY Gauge Theories and the Wilson Operator Expansion}",
    reportNumber = "IYF-86-62",
    doi = "10.1016/0550-3213(86)90451-7",
    journal = "Nucl. Phys. B",
    volume = "277",
    pages = "456",
    year = "1986"
}

@article{Kolda:1998wq,
    author = "Kolda, Christopher F. and Lyth, David H.",
    title = "{Quintessential difficulties}",
    eprint = "hep-ph/9811375",
    archivePrefix = "arXiv",
    reportNumber = "LBL-42536, LBNL-42536, LANCS-TH-9821",
    doi = "10.1016/S0370-2693(99)00657-7",
    journal = "Phys. Lett. B",
    volume = "458",
    pages = "197--201",
    year = "1999"
}

@article{Planck:2018vyg,
    author = "Aghanim, N. and others",
    collaboration = "Planck",
    title = "{Planck 2018 results. VI. Cosmological parameters}",
    eprint = "1807.06209",
    archivePrefix = "arXiv",
    primaryClass = "astro-ph.CO",
    doi = "10.1051/0004-6361/201833910",
    journal = "Astron. Astrophys.",
    volume = "641",
    pages = "A6",
    year = "2020",
    note = "[Erratum: Astron.Astrophys. 652, C4 (2021)]"
}

@article{Baer:2012up,
    author= {Baer, Howard and Barger, Vernon and Huang, Peisi
                     and Mickelson, Dan and Mustafayev, Azar and Tata, Xerxes},
    title= {Radiative Natural Supersymmetry: Reconciling
                     Electroweak Fine-Tuning and the Higgs Boson Mass},
    journal= {Phys. Rev. D},
    volume= {87},
    pages= {115028},
    year= {2013},
    doi= {10.1103/PhysRevD.87.115028},
    eprint= {1212.2655},
    archivePrefix = {arXiv},
    primaryClass = {hep-ph}
}

@article{Hebecker:2019,
    author        = "Hebecker, Arthur and Skrzypek, Torben and Wittner, Manuel",
    title         = "{The F-term problem and other challenges of stringy quintessence}",
    journal       = "JHEP",
    volume        = "11",
    pages         = "134",
    year          = "2019",
    doi           = "10.1007/JHEP11(2019)134",
    eprint        = "1909.08625",
    archivePrefix = "arXiv",
    primaryClass  = "hep-th"
}

@article{Brax:2017bcp,
    author = "Brax, Philippe and Valageas, Patrick and Vanhove, Pierre",
    title = "{$R^2$ dark energy in the laboratory}",
    eprint = "1711.03356",
    archivePrefix = "arXiv",
    primaryClass = "astro-ph.CO",
    doi = "10.1103/PhysRevD.97.103508",
    journal = "Phys. Rev. D",
    volume = "97",
    number = "10",
    pages = "103508",
    year = "2018"
}

@article{Brax:2023nvx,
    author = "Brax, Philippe and Vanhove, Pierre",
    title = "{Scalaron dynamics from the UV to the IR regime}",
    eprint = "2309.12087",
    archivePrefix = "arXiv",
    primaryClass = "gr-qc",
    reportNumber = "IPHT-t23/049, LAPTH-033/23",
    doi = "10.1103/PhysRevD.109.064088",
    journal = "Phys. Rev. D",
    volume = "109",
    number = "6",
    pages = "064088",
    year = "2024"
}

@article{Brax:2009kd,
    author = "Brax, Philippe and van de Bruck, Carsten and Martin, Jerome and Davis, Anne-Christine",
    title = "{Decoupling Dark Energy from Matter}",
    eprint = "0904.3471",
    archivePrefix = "arXiv",
    primaryClass = "hep-th",
    doi = "10.1088/1475-7516/2009/09/032",
    journal = "JCAP",
    volume = "09",
    pages = "032",
    year = "2009"
}

@article{Damour:1994zq,
    author = "Damour, T. and Polyakov, Alexander M.",
    title = "{The String dilaton and a least coupling principle}",
    eprint = "hep-th/9401069",
    archivePrefix = "arXiv",
    doi = "10.1016/0550-3213(94)90143-0",
    journal = "Nucl. Phys. B",
    volume = "423",
    pages = "532--558",
    year = "1994"
}

@article{Berge:2017ovy,
    author = "Berg{\'e}, Joel and Brax, Philippe and M{\'e}tris, Gilles and Pernot-Borr{\`a}s, Martin and Touboul, Pierre and Uzan, Jean-Philippe",
    title = "{MICROSCOPE Mission: First Constraints on the Violation of the Weak Equivalence Principle by a Light Scalar Dilaton}",
    eprint = "1712.00483",
    archivePrefix = "arXiv",
    primaryClass = "gr-qc",
    doi = "10.1103/PhysRevLett.120.141101",
    journal = "Phys. Rev. Lett.",
    volume = "120",
    number = "14",
    pages = "141101",
    year = "2018"
}

@article{Hamaguchi:2018vtv,
    author = "Hamaguchi, Koichi and Ibe, Masahiro and Moroi, Takeo",
    title = "{The swampland conjecture and the Higgs expectation value}",
    eprint = "1810.02095",
    archivePrefix = "arXiv",
    primaryClass = "hep-th",
    reportNumber = "UT-18-19, IPMU18-0159",
    doi = "10.1007/JHEP12(2018)023",
    journal = "JHEP",
    volume = "12",
    pages = "023",
    year = "2018"
}

@article{Malquarti:2002bh,
    author = "Malquarti, Michael and Liddle, Andrew R",
    title = "{Initial conditions for quintessence after inflation}",
    eprint = "astro-ph/0203232",
    archivePrefix = "arXiv",
    doi = "10.1103/PhysRevD.66.023524",
    journal = "Phys. Rev. D",
    volume = "66",
    pages = "023524",
    year = "2002"
}

@article{Martin:2019lqd,
    author = "Martin, Stephen P. and Robertson, David G.",
    title = "{Standard model parameters in the tadpole-free pure $\overline{\rm{MS}}$ scheme}",
    eprint = "1907.02500",
    archivePrefix = "arXiv",
    primaryClass = "hep-ph",
    doi = "10.1103/PhysRevD.100.073004",
    journal = "Phys. Rev. D",
    volume = "100",
    number = "7",
    pages = "073004",
    year = "2019"
}

@article{Garriga:2003hj,
    author = "Garriga, Jaume and Linde, Andrei D. and Vilenkin, Alexander",
    title = "{Dark energy equation of state and anthropic selection}",
    eprint = "hep-th/0310034",
    archivePrefix = "arXiv",
    doi = "10.1103/PhysRevD.69.063521",
    journal = "Phys. Rev. D",
    volume = "69",
    pages = "063521",
    year = "2004"
}

\end{document}